\pdfoutput=1
\documentclass[twocolumn]{aastex62}

\usepackage{amsmath}
\usepackage{graphicx}
\usepackage{url}
\usepackage{tabularx}
\usepackage{footmisc}
\usepackage{epstopdf}
\usepackage{hyperref}
\usepackage{breakurl}
\usepackage{rotating}
\usepackage{longtable}
\usepackage{multirow}
\usepackage{ctable}
\newcommand{\etaEarth}{\ensuremath{\eta_{\oplus}}}

\expandafter\def\expandafter\UrlBreaks\expandafter{\UrlBreaks
  \do\a\do\b\do\c\do\d\do\e\do\f\do\g\do\h\do\i\do\j%
  \do\k\do\l\do\m\do\n\do\o\do\p\do\q\do\r\do\s\do\t%
  \do\u\do\v\do\w\do\x\do\y\do\z\do\A\do\B\do\C\do\D%
  \do\E\do\F\do\G\do\H\do\I\do\J\do\K\do\L\do\M\do\N%
  \do\O\do\P\do\Q\do\R\do\S\do\T\do\U\do\V\do\W\do\X%
  \do\Y\do\Z}

\newcommand{\kepler}{\emph{Kepler}}

\published{in AAS Journals}

\shorttitle{DR25 Planet Candidate catalog Completeness}
\shortauthors{Christiansen et al.}

\begin{document}


\title{Measuring Transit Signal Recovery in the Kepler Pipeline IV: Completeness of the DR25 Planet Candidate catalog}


\correspondingauthor{J. L. Christiansen}
\email{christia@ipac.caltech.edu}

\author[0000-0002-8035-4778]{Jessie L. Christiansen}
\affiliation{IPAC, Mail Code 100-22, Caltech, 1200 E. California Blvd. Pasadena CA 91125}

\author{Bruce D. Clarke}
\affiliation{SETI Institute, 189 Bernardo Ave, Suite 200, Mountain View, CA 94043, USA}
\affiliation{NASA Ames Research Center, Moffett Field, CA 94035}

\author[0000-0002-7754-9486]{Christopher~J.~Burke}
\affiliation{Kavli Institute for Astrophysics and Space Research, Massachusetts Institute of Technology, Cambridge, MA}

\author{Jon M. Jenkins}
\affiliation{NASA Ames Research Center, Moffett Field, CA 94035}

\author[0000-0003-0081-1797]{Stephen T. Bryson}
\affiliation{NASA Ames Research Center, Moffett Field, CA 94035}

\author[0000-0003-1634-9672]{Jeffrey L. Coughlin}
\affiliation{SETI Institute, 189 Bernardo Ave, Suite 200, Mountain View, CA 94043, USA}
\affiliation{NASA Ames Research Center, Moffett Field, CA 94035}

\author[0000-0001-7106-4683]{Susan E. Mullally}
\affiliation{Space Telescope Science Institute, 3700 San Martin Dr, Baltimore, MD 21218}

\author{Joseph D. Twicken}
\affiliation{NASA Ames Research Center, Moffett Field, CA 94035}
\affiliation{SETI Institute, 189 Bernardo Ave, Suite 200, Mountain View, CA 94043, USA}

\author{Natalie M. Batalha}
\affiliation{University of California Santa Cruz, Santa Cruz, CA}

\author{Joseph Catanzarite}
\affiliation{NASA Ames Research Center, Moffett Field, CA 94035}
\affiliation{SETI Institute, 189 Bernardo Ave, Suite 200, Mountain View, CA 94043, USA}

\author{AKM Kamal Uddin}
\affiliation{NASA Ames Research Center, Moffett Field, CA 94035}
\affiliation{SETI Institute, 189 Bernardo Ave, Suite 200, Mountain View, CA 94043, USA}

\author{Khadeejah Zamudio}
\affiliation{formerly NASA Ames Research Center, Moffett Field, CA 94035}
\affiliation{formerly KBRwyle, Houston, TX}

\author{Jeffrey C. Smith}
\affiliation{NASA Ames Research Center, Moffett Field, CA 94035}
\affiliation{SETI Institute, 189 Bernardo Ave, Suite 200, Mountain View, CA 94043, USA}

\author{Christopher E. Henze}
\affiliation{NASA Ames Research Center, Moffett Field, CA 94035}

\author{Jennifer Campbell}
\affiliation{formerly NASA Ames Research Center, Moffett Field, CA 94035}
\affiliation{formerly KBRwyle, Houston, TX}



\begin{abstract}

In this work we empirically measure the detection efficiency of \kepler\ pipeline used to create the final \kepler\ Threshold Crossing Event \cite[TCE;][]{Twicken2016} and planet candidate catalogs \citep{Thompson2018}, a necessary ingredient for occurrence rate calculations using these lists. By injecting simulated signals into the calibrated pixel data and processing those pixels through the pipeline as normal, we quantify the detection probability of signals as a function of their signal strength and orbital period. In addition we investigate the dependence of the detection efficiency on parameters of the target stars and their location in the \kepler\ field of view. We find that the end-of-mission version of the \kepler\ pipeline returns to a high overall detection efficiency, averaging a 90--95\% rate of detection for strong signals across a wide variety of parameter space. We find a weak dependence of the detection efficiency on the number of transits contributing to the signal and the orbital period of the signal, and a stronger dependence on the stellar effective temperature and correlated noise properties. We also find a weak dependence of the detection efficiency on the position within the field of view. By restricting the \kepler\ stellar sample to stars with well-behaved correlated noise properties, we can define a set of stars with high detection efficiency for future occurrence rate calculations.

\end{abstract}





\section{Introduction}

The Data Release 25 (DR25) planet candidate catalog from the NASA \emph{Kepler} mission \citep{Thompson2018} represents the culmination of eight years' worth of analysis. The list of 4,034 planet candidates was generated in a fully automated fashion from the full \emph{Kepler} observation baseline of nearly four years. Using an automatic classification scheme called the Robovetter, the Threshold Crossing Events (TCEs) generated by the \emph{Kepler} data reduction pipeline were dispositioned as either planet candidates or false positives. This automation allowed, for the first time, an attempt to quantify the completeness (false negative rate) and reliability (false positive rate) of the catalog. These comprise two of the necessary ingredients for measuring the underlying planet occurrence rates from an observed list of planet candidates.

We have run a series of experiments characterising the completeness of almost all the recent versions of the \emph{Kepler} pipeline, increasing in scope and complexity with each iteration. The work presented here represents the analysis of the fourth such experiment. Previous results can be found in \citet[][SOC version 9.2, spanning the full observing baseline]{Christiansen2016}, \citet[][SOC version 9.1, spanning one year of observations]{Christiansen2015}, and \citet[][an early version of SOC version 8.3, spanning one month of observations]{Christiansen2013}. Each version of the pipeline has its own strengths and weaknesses, and the measured completeness of the accompanying planet candidate catalogs shows significant variations in each case. Therefore, it is crucial for studies of exoplanet demographics that the completeness model used to analyse a given planet candidate catalog be derived from the corresponding pipeline version. The work presented here and in \citet{Christiansen2017} quantifies the completeness of the pipeline version used to generate the catalog published by \citet{Thompson2018}; the corresponding quantification of the Robovetter is presented in \citet{Coughlin2017}. \cite{Christiansen2016} accompanies the catalog published by \citet{Coughlin2016}; \cite{Christiansen2015} accompanies the catalog published by \citet{Mullally2015}; and \citet{Christiansen2013} corresponds to an early version of the code used to produce the catalog published by \citet{Rowe2015}. In all previous cases there are important caveats in the careful usage and application of the measured completeness; readers should refer to the relevant citations for additional details. In addition, we note that this work does not quantify the false positive rate nor false alarm rate of the \kepler\ pipeline; these pieces must be calculated and included separately in occurrence rate calculations.

In \citet{Christiansen2017} and \citet{Thompson2018}, we presented an initial analysis of the completeness of the DR25 planet candidate catalog to support calculations of $\etaEarth$, the frequency of Earth-like planets orbiting stars like the Sun. This early analysis was restricted to producing a single 1-dimensional measure of the detection efficiency as a function of the Multiple Event Statistic (MES) of the transit signal, i.e. its signal strength, for all FGK dwarf stars in the \emph{Kepler} target list. In this paper we extend that analysis to investigate the completeness of the \kepler\ pipeline along several additional axes, for use in defining and supporting additional science use cases. The paper is organized as follows: in Section \ref{sec:pipeline}, we describe the generation of the DR25 planet candidate catalog. In Section \ref{sec:experiment}, we describe the details and execution of this fourth transit injection experiment, and in Section \ref{sec:results} we explore the results. In Section \ref{sec:discussion} we discuss the implications, and in Section \ref{sec:conclusion} we summarise the final results.

\section{DR25 Planet Candidate Catalog Generation}
\label{sec:pipeline}

The DR25 planet candidate catalog was produced by uniformly processing the full, four-year dataset (Quarters 1 through 17) through Science Operations Center (SOC) pipeline version 9.3 to produce a list of Threshold Crossing Events (TCEs), described by \citet{Twicken2016}. The TCEs were then evaluated by the Robovetter \citep{Coughlin2016,Thompson2018}, and dispositioned into planet candidates or false positives. In order to empirically recover the detection efficiency of the process---the likelihood a given planet signal will be correctly identified and dispositioned as a planet candidate---we can replicate the process with a suite of known `ground-truth' signals, and analyse their outcomes. We summarise the pipeline and Robovetter processes in the Appendix, with particular emphasis on updates compared to the previous planet candidate catalogs. 

\section{Pixel-Level Transit Injection}
\label{sec:experiment}

In order to characterize the pipeline detection efficiency, we have performed several distinct transit injection experiments. These largely fall into two categories: pixel-level transit injection (PLTI) and flux-level transit injection (FLTI). For PLTI experiments, simulated transit signals are injected into the calibrated pixels, before the aperture photometry time series is constructed and cotrended. This allows the total detection efficiency loss to be determined through the photometric and search portions of the pipeline. However, PLTI is computationally expensive, since it run 
most of the pipeline modules. As a result, these PLTI experiments are limited to one injected planetary signal per target star, but include all available target stars. Hence, PLTI provides an average detection efficiency over a set of stars. Knowing that the stars are not all ‘average’, a series of FLTI experiments were also conducted. For FLTI, the transit signal is injected into the cotrended flux time series within the Transiting Planet Search (TPS) module of the pipeline, and the signal detection algorithm is performed over a restricted portion of the period search space focused on the period of the injected signal \citep{BurkeJCat2017b}. For ‘deep’ FLTI experiments, we chose a small subset ($\sim$100) of stars and performed $\sim$600,000 injection and recovery experiments for each star. For ‘shallow’ FLTI experiments, we chose a larger subset ($\sim$30,000) of stars and performed $\sim$2,000 injection and recovery experiments for each star. These tests determined when and how individual stars can deviate from the average detection efficiency measured by PLTI. This document describes the PLTI experiment only; the FLTI products are documented separately in \citet{BurkeJCat2017b} and examples of using FLTI products to measure detection efficiency are discussed in \citet{BurkeJCat2017a}.


For the PLTI experiment described here, we inject simulated transit and eclipse signals into the calibrated pixels of 190,128 targets covering the entire focal plane. These injections fell into three distinct categories, each designed for a specific use case: 146,295 targets were injected with planet-like signals at the target location on the CCD, thereby mimicking a planet orbiting the specified target; 33,978 targets were injected with planet-like signals at a location slightly offset from the nominal target location, thereby mimicking a blended eclipsing binary; and 9,856 targets were injected with eclipsing-binary-like signals (having both primary and secondary eclipses) at the target location on the CCD. The latter two groups were generated to test the Robovetter's ability to discriminate between various kinds of false positives \citep[for a detailed analysis, see][]{Coughlin2017}, and are made available along with the first group for the community to test their own algorithms. The analysis described in this work will be restricted to the on-target planet-like signals in the first group, and to the completeness of the \kepler\ pipeline, not the subsequent Robovetter stage. To generate the simulated transit signals we use the DR25 Q1--Q17 stellar parameters provided by \citet{Mathur2017ApJS}. An updated set of stellar parameters was released by \citet{berger18} during this analysis, but was not found to systematically change the conclusions.

\begin{figure*}[ht!]
\centering 
\includegraphics[width=16cm]{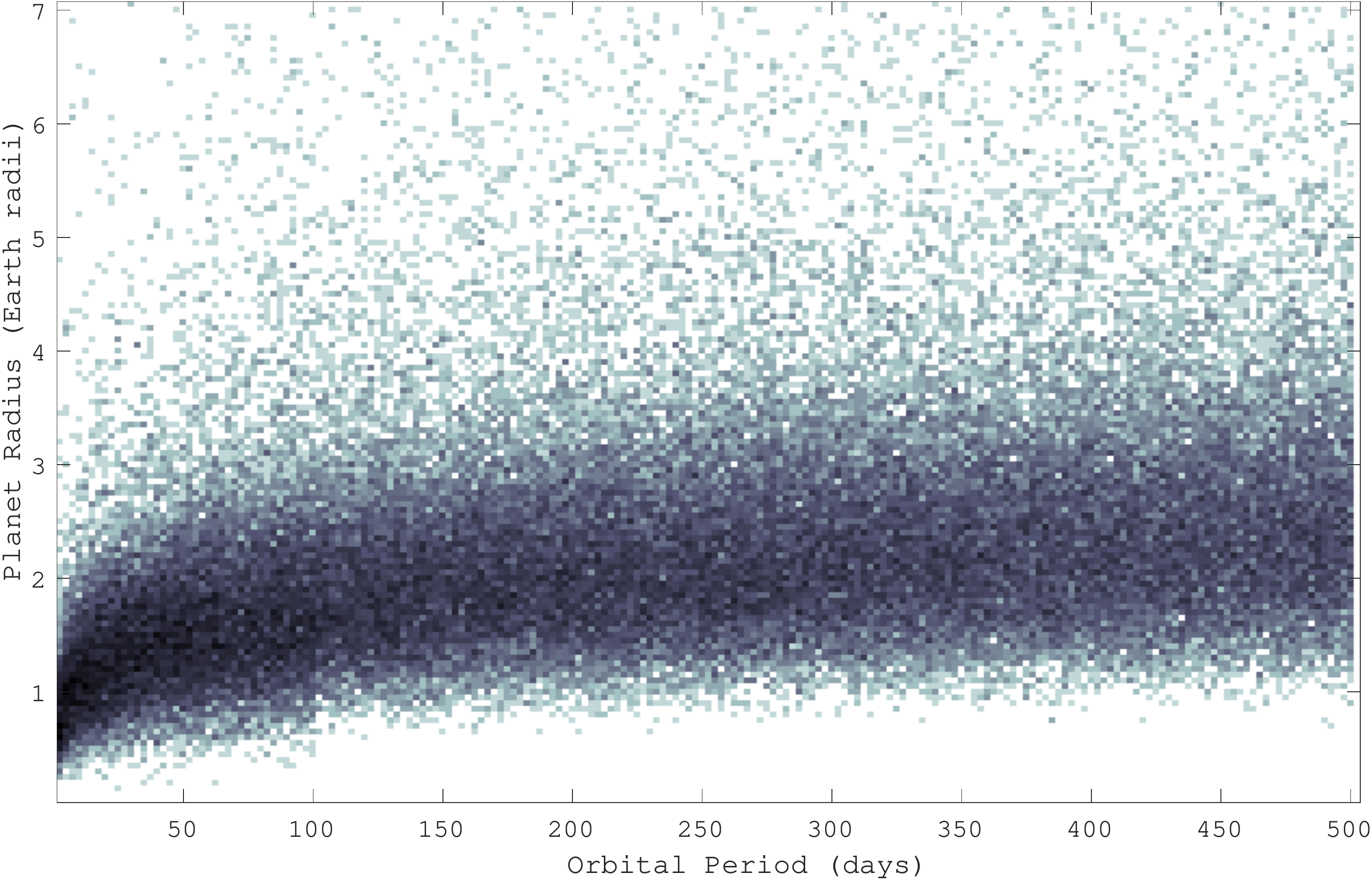}
\caption{A density distribution of simulated planet signals injected on-target, in planet radius and orbital period space (truncated at 7$R_{\oplus}$). The injections are clustered around the pipeline signal-to-noise threshold, which moves to larger radii for longer periods, in order to examine the transition from detection to non-detection by the pipeline.}
\label{fig:injectionrecoveries}
\end{figure*}

\begin{figure}
\centering 
\includegraphics[width=8.5cm]{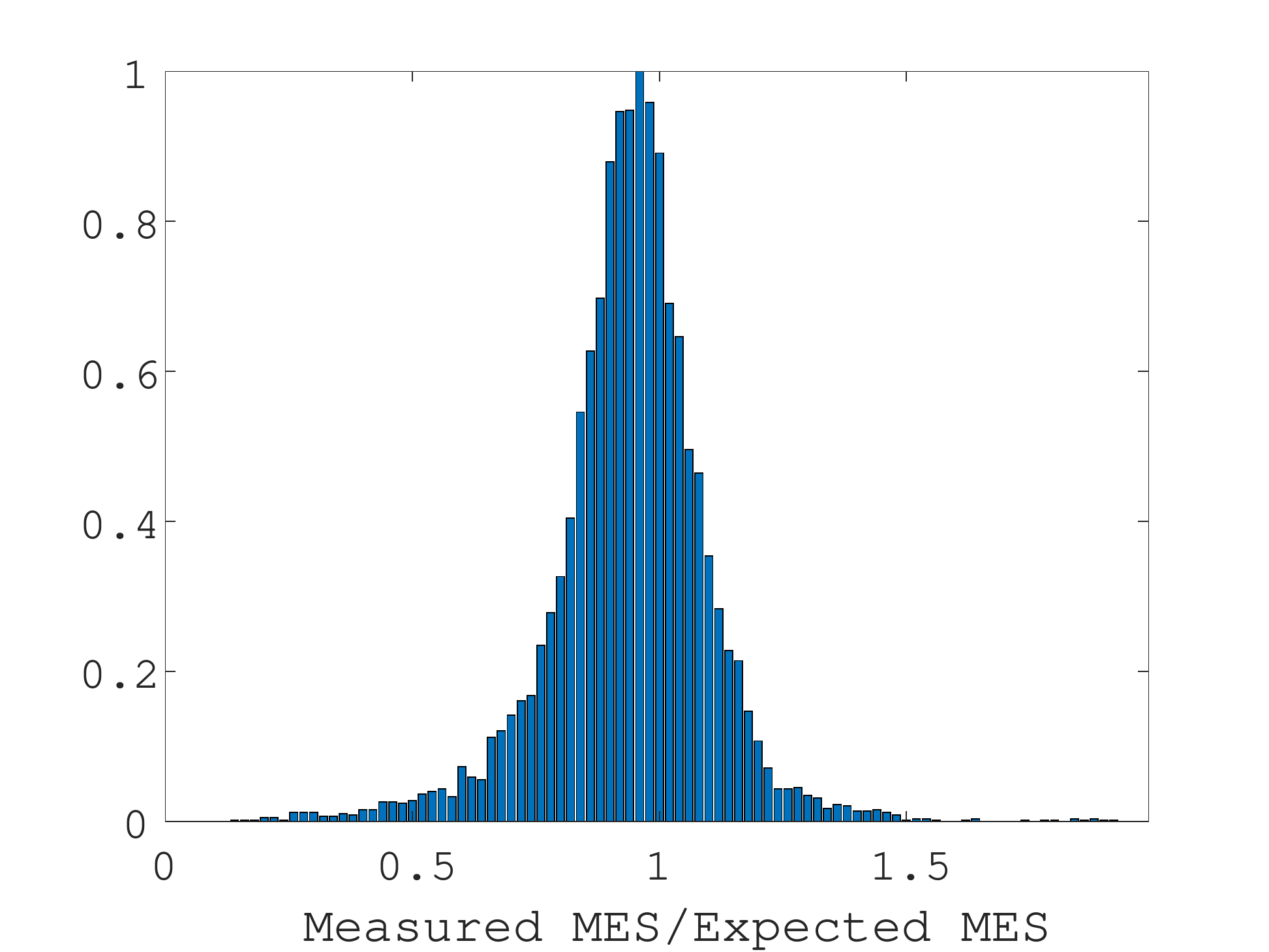}
\caption{{\bf A histogram of the values of the measured MES of the successfully recovered injected signals compared to the estimated MES of the signal prior to injection.}}
\label{fig:mescomparison}
\end{figure}

For non-M-dwarf targets, each injected signal was generated as follows. First, the orbital period was drawn from a uniform range in period spanning 0.5--500 days. The planet radius was then chosen such that the resulting MES spanned the range 0--20, bracketing the pipeline transition from fully complete (100\% signal recovery) to fully incomplete (0\% recovery). To estimate the MES for a given injected signal we take into account the stellar radius, orbital period, planet radius, an average Combined Differential Photometric Precision \citep[rmsCDPP;][]{JenkinsSPIE2010,Christiansen2012}, the dilution of the signal by additional light in the photometric aperture, and the duty cycle of the observations (discarding gapped cadences and de-weighted cadences with weights $<0.5$. We note that this process resulted in some unphysically large radii, but ultimately 50\% of the injected planets have radii $<2R_{\oplus}$ and 90\% have radii $<40R_{\oplus}$. The orbital eccentricity was fixed at 0, and the impact parameter was drawn from a uniform range spanning 0--1. We also note that, in general, the MES that we estimate for the signal prior to its injection in the light curve will not equal the measured MES. We estimate the MES using a single rmsCDPP value, which is the average noise over the light curve---the actual data points into which the signal is injected may have higher or lower levels of local noise. Figure \ref{fig:mescomparison} shows the value of the measured MES compared to the estimated MES; the median value is 0.95, with a standard deviation of 0.15. The overall reduction in the measured MES is at least partly due to the average timing mismatch between the injected period, epoch, and transit duration and the finite grid of periods, epochs, and transit durations over which the pipeline searches \cite[see, e.g][]{Jenkins1996,Jenkins2010}.

For the M-dwarf targets, where the habitable zone is much closer to the star, we concentrated the injected signals over a smaller period range. For the 3,809 targets with $2400 K \le T_{\rm eff} \le 3900 K$ and $log \rm{g} \ge 4$, the orbital periods were selected from a uniform range spanning 0.5--100 days. Given the stellar radii, noise properties, and orbital periods, this resulted in a smaller injected planet radius distribution than the remainder of the targets, with 50\% of the injected planets having radii $<0.92R_{\oplus}$ and 90\% have radii $<1.7R_{\oplus}$.

\section{Results}
\label{sec:results}

The full table of parameters for the injected signals, and if they were recovered, the parameters of the recovered signal, is available at the NASA Exoplanet Archive\footnote{\url{https://exoplanetarchive.ipac.caltech.edu/docs/KeplerSimulated.html}, DOI 10.26133/NEA14}. All three types of injected signal are included (planet injected on-target, planet injected off-target, and EB injected on-target); for the remainder of this section we focus solely on analysis of the on-target planet signal injections. Of the 146,295 planets, 45,281 were successfully recovered by the pipeline, as shown in Figure \ref{fig:injectionrecoveries}. We note here that many signals were injected below the signal detection threshold for the express purpose of exploring the transition region from detection to non-detection. For these purposes, `success' is defined by the ephemeris-matching algorithm described in Section 6.2 of \citet{Thompson2018}. This incorporates both a period tolerance and a check on the number of transit events, and allows for detections with periods that differ by half/double or a third/thrice the injected period. In the remainder of this Section we explore the detection efficiency of the pipeline with respect to several variables.

\subsection{Orbital Period}
\label{sec:orbitalperiod}

\begin{figure}[ht!]
\centering 
\includegraphics[width=8cm]{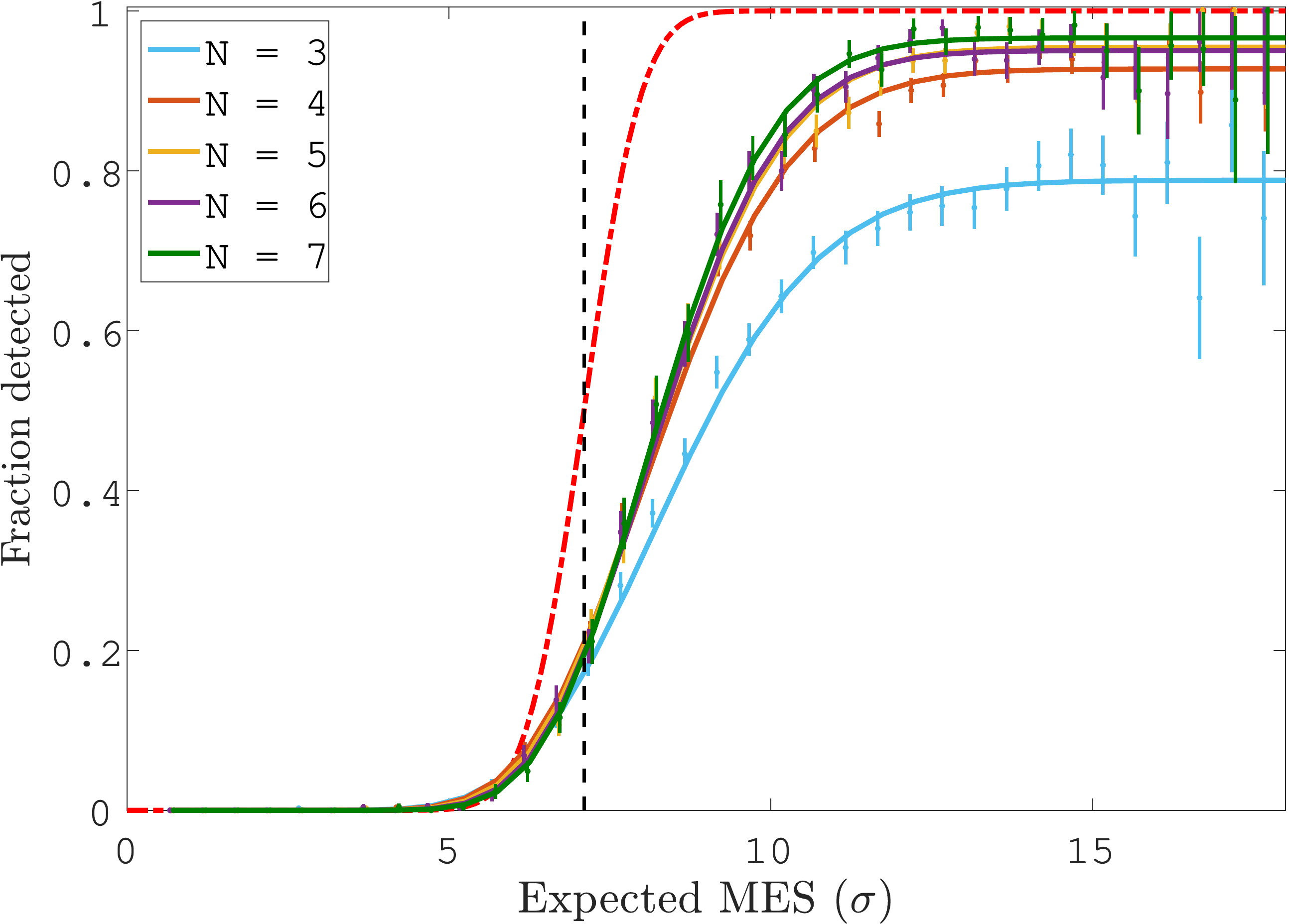}
\includegraphics[width=8cm]{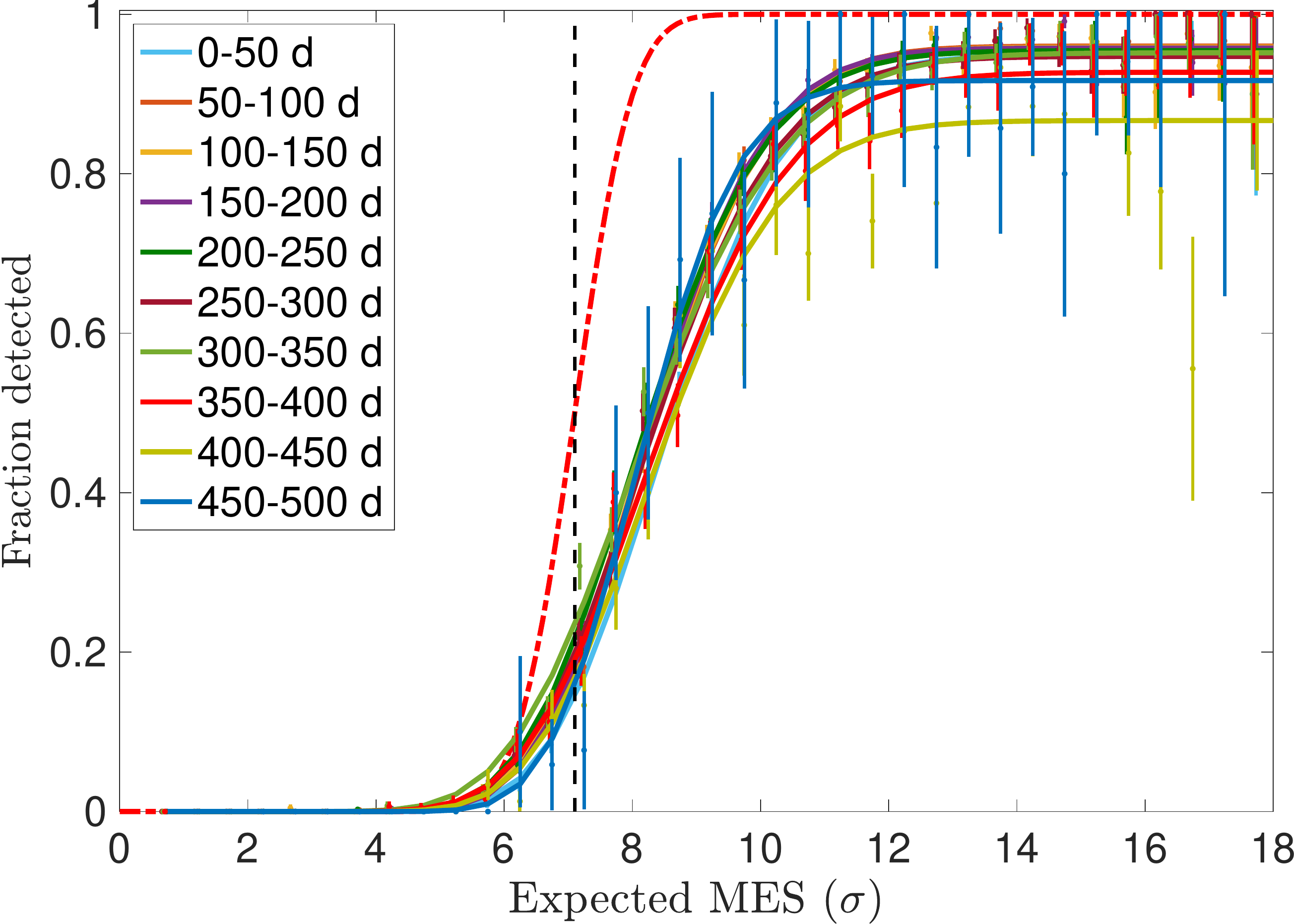}
\caption{{\bf Upper:} The 1-dimensional detection efficiency of the pipeline calculated in different numbers of transits ($N$) contributing to the MES. The curves are $\Gamma$ cumulative distribution functions fit to the data. The binomial uncertainties on the data in each bin are shown with small horizontal offsets for clarity. The black dashed line shows the 7.1-$\sigma$ detection threshold of the pipeline, and the red dashed line shows the hypothetical perfect performance of the pipeline for pure white noise. {\bf Middle:} As above, but calculated for different orbital period ranges. {\bf Lower:} As above, but calculated for different ranges of stellar effective temperature.}
\label{fig:perioddependence}
\end{figure}

There is an expected drop in detection efficiency at longer periods due simply to the window function of the observations---beyond some period, meeting the minimum number of transits required for detection becomes less and less likely. For SOC version 9.3, there is an additional penalty applied during TPS to signals with only three contributing transits \cite[see Section 9.4.4.1 of][]{Jenkins2017b}. The top panel of Figure \ref{fig:perioddependence} shows the detection efficiency of the pipeline as a function of the number of transits contributing to the detection. For the 77,860 targets with at least four transits with durations shorter than 15 hours (the longest duration searched by the pipeline), we assess the fraction of injected signals that are recovered as a function of the number of contributing transits. As we have done in previous work \citep{Christiansen2013, Christiansen2015, Christiansen2016}, we analyse the detection efficiency in each bin as a function of the expected MES, fitting a $\Gamma$ cumulative distribution function of the form

\begin{equation}
	p = F(x|a,b)=\frac{c}{b^a\Gamma(a)}\int\limits_0^x t^{a-1}e^{-t/b}dt
	\label{eq:gamma}
\end{equation}

\noindent normalized to the value ($c$) at MES = 15. The additional penalty for having only three transits is clear, and has subsequently been incorporated in the generation of the window function \citep{Burke2017window}. For the period analysis performed here we only consider injections with four or more transits. We find a remaining dependence of the detection efficiency on orbital period, shown in the middle panel of Figure \ref{fig:perioddependence}. From 0--300 days there is little change in the overall detection efficiency, barring a small drop in sensitivity in the 0--50 day bin due to the known behavior of the harmonic fitter removing signals at short periods \citep{Christiansen2013,Christiansen2015}. For periods longer than 300 days the detection efficiency falls off slightly, from 95--96\% at the shorter periods to 87--91\% at the longest periods. This is an important caveat for those calculating occurrence rate in this interesting long-period parameter space. Given the dependence of the detection efficiency on orbital period for periods longer than 300 days, we recommend determining the detection efficiency over the period range of interest for a given occurrence rate calculation, rather than relying on the ensemble average detection efficiency, and to pay particular attention to the window function.

\subsection{Stellar Properties}
\label{sec:stellartype}

\begin{figure*}
  \includegraphics[width=0.48\textwidth]{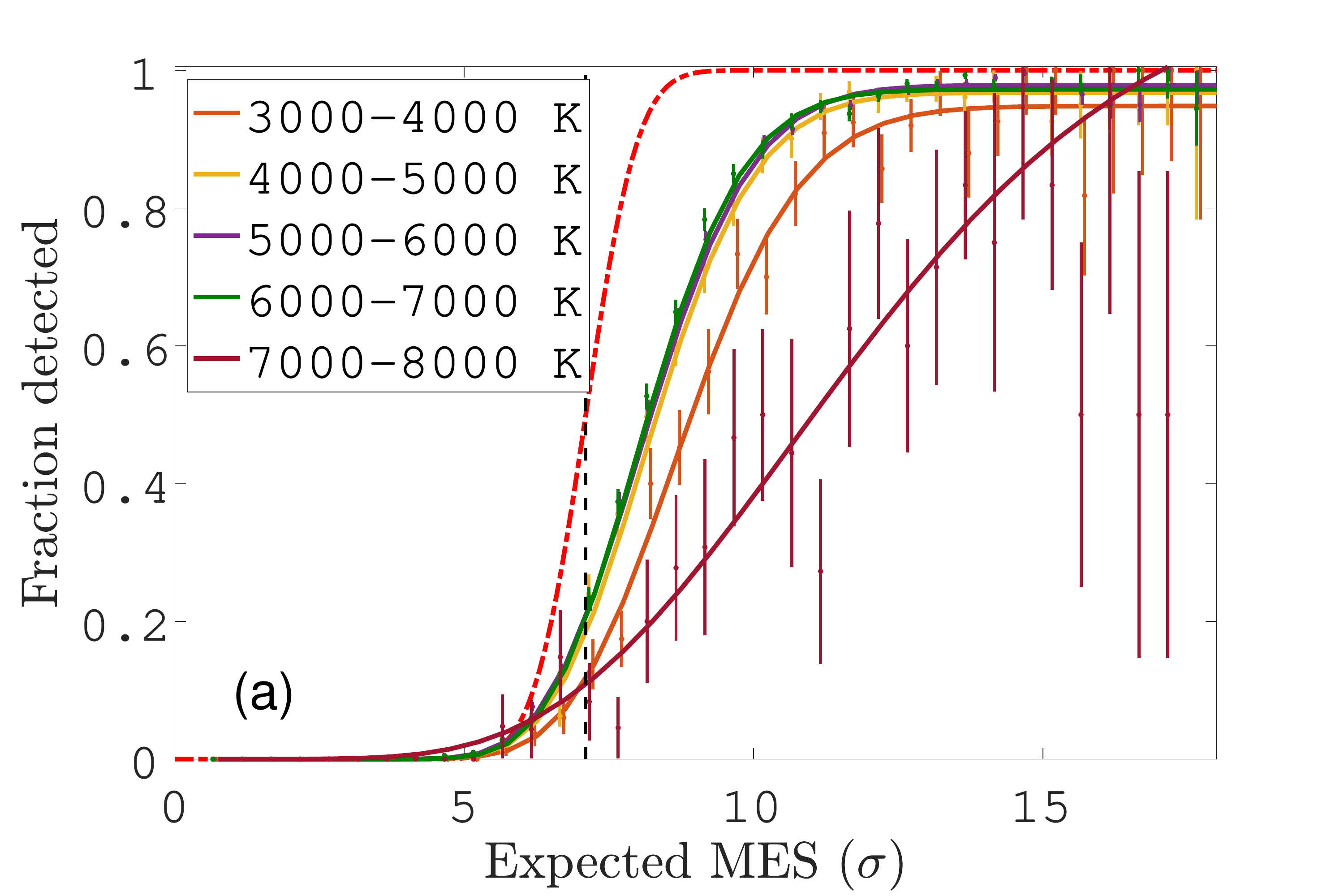}
  \includegraphics[width=0.48\textwidth]{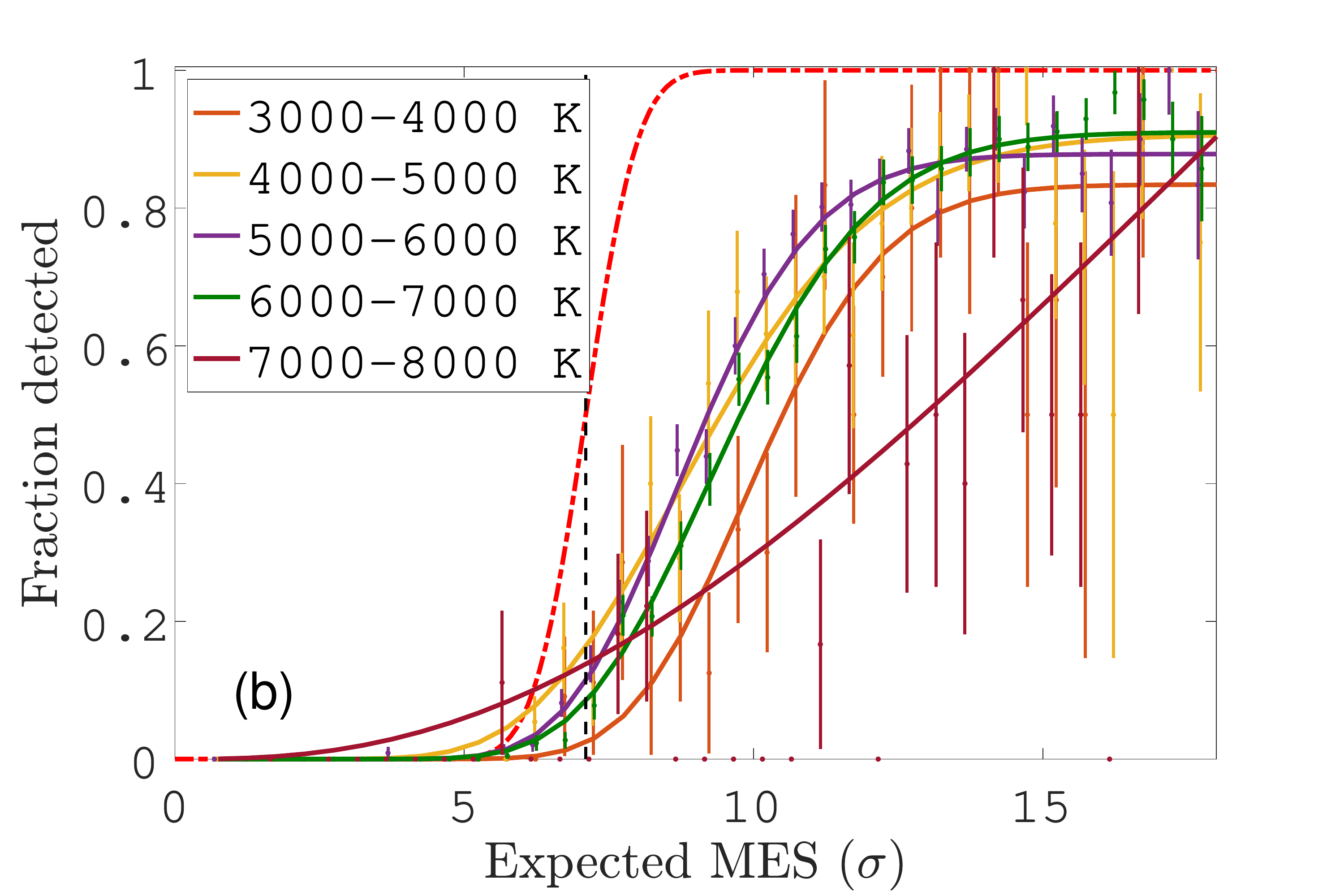}
  \caption{As for Figure \ref{fig:perioddependence}, but calculated for different ranges in stellar effective temperature. {\it Left}: Stars with low (CDPP$_{\rm L}<-0.2$. {\it Right}: Stars with high (CDPP$_{\rm L}>0.0$.)}
    \label{fig:teffcomparison}
\end{figure*}

\begin{figure*}
  \includegraphics[width=0.48\textwidth]{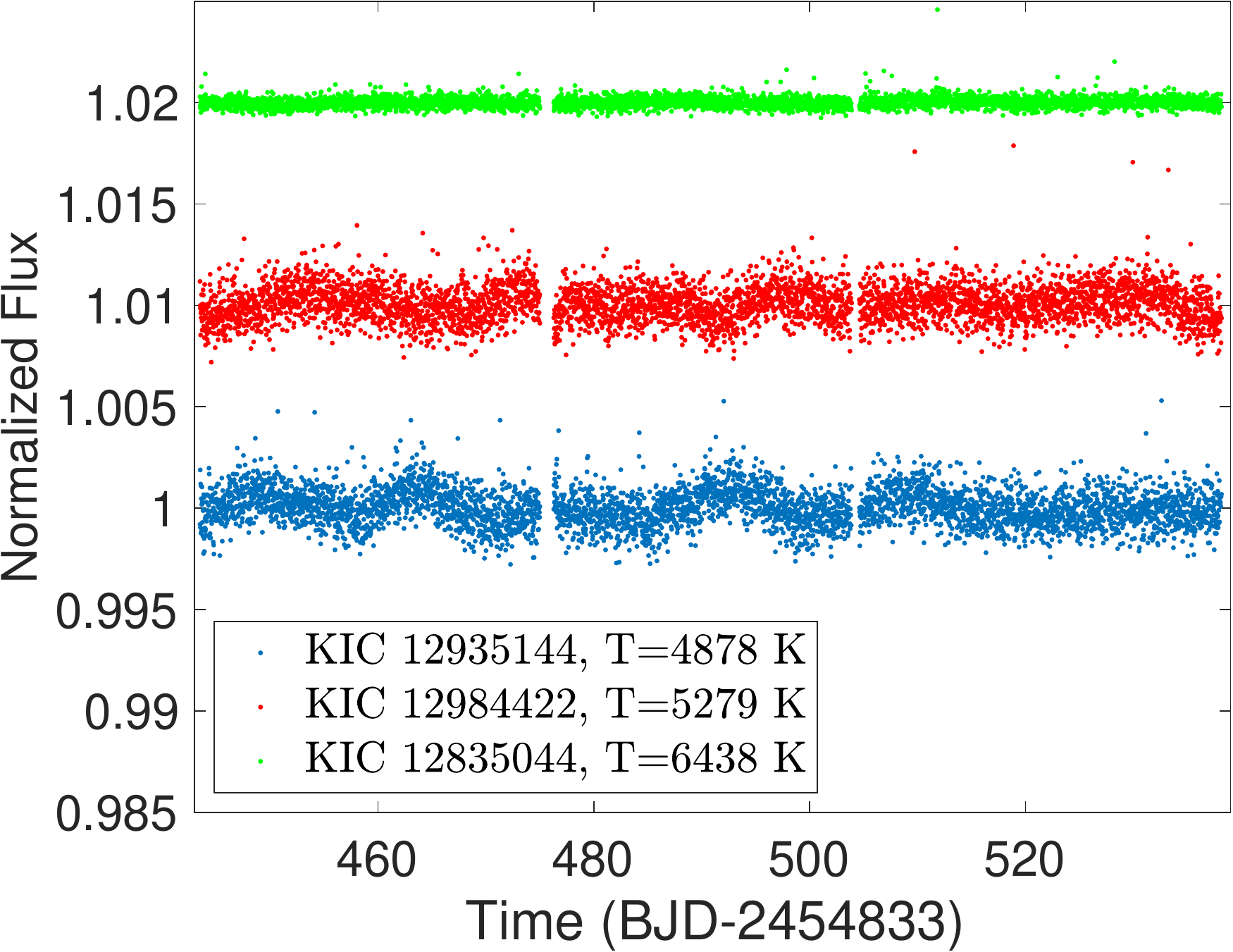}
  \includegraphics[width=0.48\textwidth]{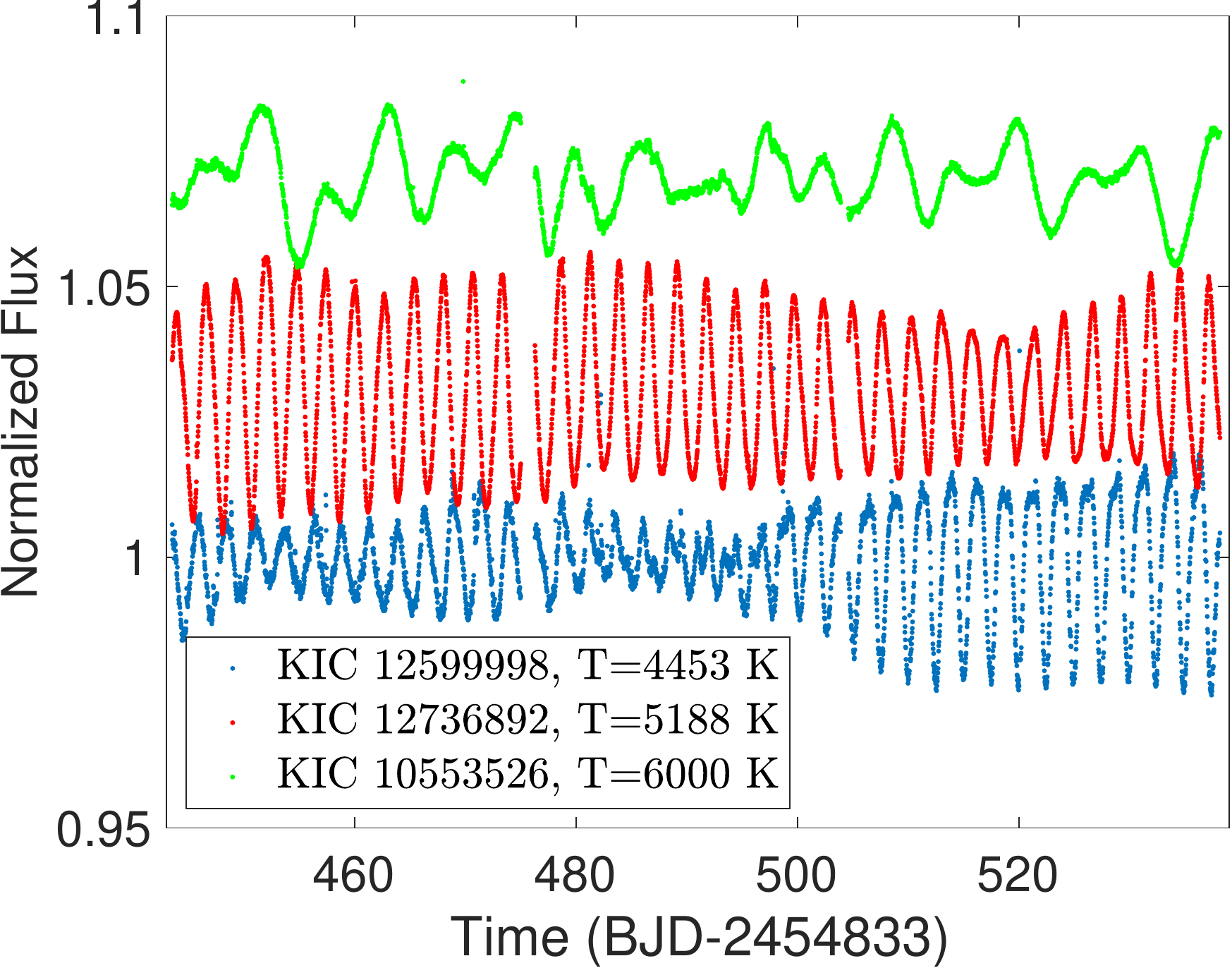}
  \caption{{\it Left}: Quarter 5 light curves of a selection of targets with CDPP$_{\rm L}<-0.3$. {\it Right}: The same for targets with CDPP$_{\rm L}>0.1$. Note the difference in the y axes.}
    \label{fig:cdppslopecomparison}
\end{figure*}

The {\it Kepler} pipeline measures the noise in a given time series using the Combined Differential Photometric Precision \citep[CDPP;][]{Christiansen2012}. The CDPP is calculated at each data point in the time series for a set of 14 different trial transit durations, and is equivalent to the effective white noise seen by a transit pulse of that duration. An average root-mean-square CDPP (rmsCDPP) value is calculated for each time series and transit duration. For a time series that is dominated by white noise, we expect the rmsCDPP to decrease with increasing transit duration, i.e. for a transit duration that is twice as long, and integrating over twice as many data points, we expect the rmsCDPP to be reduced by a factor of $\sqrt{2}$. Therefore, we can use the change in the rmsCDPP as a function of increasing transit duration to track how well the noise for a given time series is approximated by white noise. We refer to this as the CDPP slope \citep[see Section 3.1 of][for details]{BurkeJCat2017b}; it is calculated for short (2--4.5 hr) and long (7.5--15 hr) transit durations, and provided as part of the {\it Kepler} stellar parameters at the NASA Exoplanet Archive \citep{Akeson2013}. A simulated transit signal will reach the same SNR in two different light curves if they have the same CDPP. However, if they have different CDPP slopes the ability of the pipeline to extract the transit signal from within the correlated noise is impacted. \citet{BurkeJCat2017b} found previously that the long CDPP slope (CDPP$_{\rm L}$) was the most useful discriminator, and the following analysis is based on CDPP$_{\rm L}$. CDPP$_{\rm L}$ tracks the SNR of the stellar variability amplitude on rotation period and some pulsation period time-scales. In order to visualise how CDPP$_{\rm L}$ represents the noise in the data, we show in Figure \ref{fig:cdppslopecomparison} light curves of targets with negative CDPP$_{\rm L}$ (left) and positive CDPP$_{\rm L}$ (right). The stars in the right panel have increasing rmsCDPP values with increasing trial transit durations, due to the longer transit durations encompassing a higher amplitude of intrinsic stellar variability. On the other hand, the stars in the left panel have decreasing rmsCDPP values with increasing trial transit durations, as the noise bins down as expected for predominately white noise.

Figure \ref{fig:teffcomparison} shows the detection efficiency of the pipeline broken out by stellar effective temperature for stars with low (CDPP$_{\rm L}<-0.2$, left panel) and high (CDPP$_{\rm L}>0.0$, right panel) levels of correlated noise. For this analysis, we have limited the stellar sample to stars with log$g>4.0$, to injections with four or more transits (in order to remove the confounding factor of the detection efficiency on the number of transits discussed in Section \ref{sec:orbitalperiod}), and with injected transit durations shorter than 15 hours. This leaves 91,672 targets with on-target simulated planet signals. Figure \ref{fig:teffcomparison} shows that for stars with low levels of correlated noise, those with stellar effective temperatures between 4000--7000~K (roughly FGK stars) have a well-behaved detection efficiency. For cooler dwarfs, the detection efficiency drops off slightly, plateauing at 92\% compared to $\sim$96\% for the FGK stars. For hotter stars, the detection efficiency drops off somewhat more, although the number of recovered injections (166) is too low for a robust analysis. This decline in detection efficiency for non-FGK stars was also noted in earlier versions of the pipeline \citep{Christiansen2015}. We note that the decrease in detection efficiency for the larger stars is not related to their larger stellar radii relative to the injected planet radii, as this has been accounted for when injecting the planets and calculating the expected MES. We examined the effect of using the updated Gaia DR2 stellar parameters of \citet{berger18} here and found that the stellar temperature agreement between the two sets of parameters was very high---for the range of stellar properties examined here, $T_{\rm eff,Kep} = 0.9998\pm0.0027\times T_{\rm eff,Gaia}$, where $T_{\rm eff,Kep}$ and $T_{\rm eff,Gaia}$ are the stellar effective temperature from the Kepler DR25 stellar catalog and Gaia DR2, respectively. 


\begin{figure}
\centering 
\includegraphics[width=8cm]{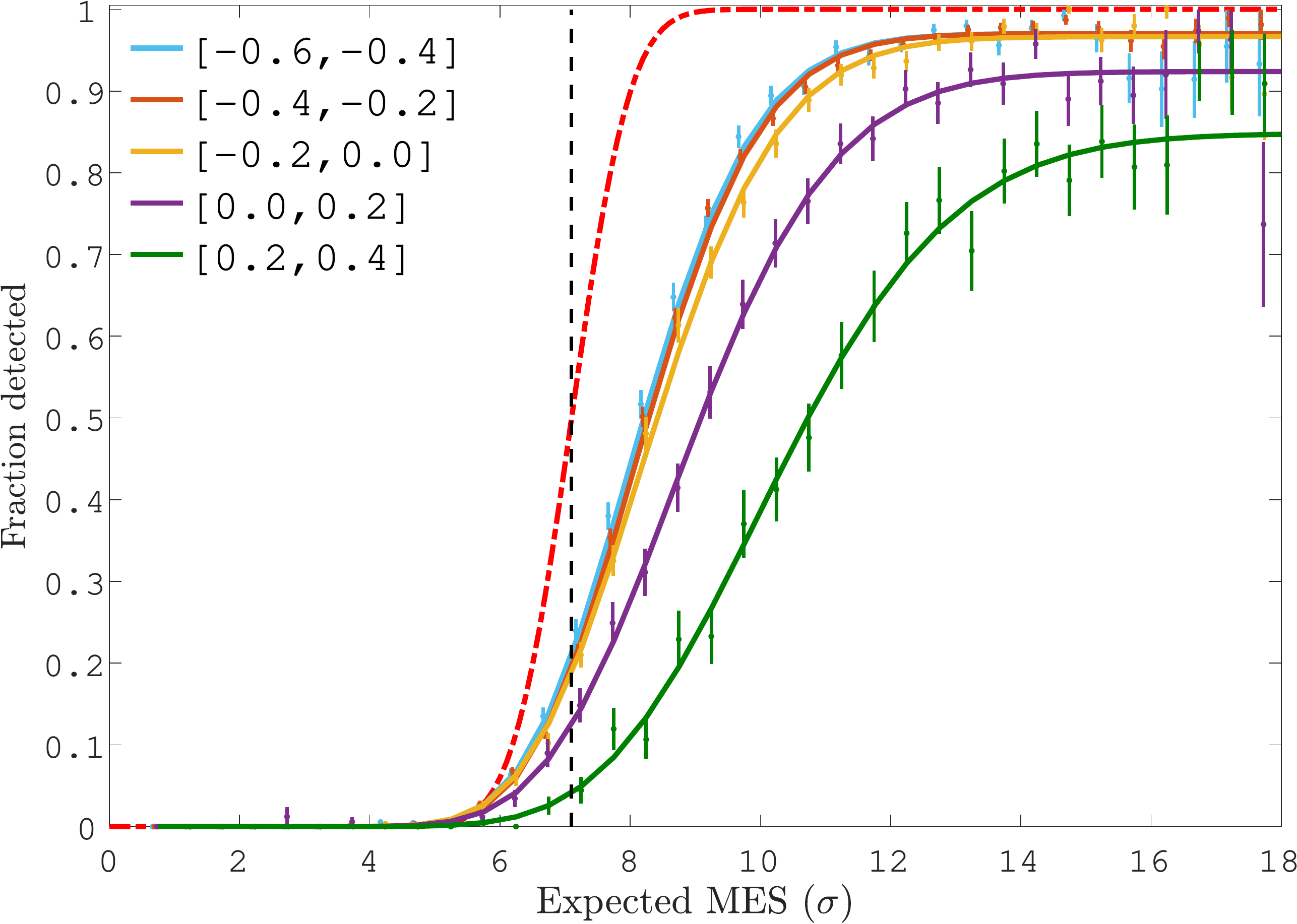}
\caption{As for Figure \ref{fig:perioddependence}, but calculated for different ranges in the long CDPP slope (CDPP$_{\rm L}$).}
\label{fig:cdppdependence}
\end{figure}

The picture is qualitatively similar but quantitatively lower for stars with high levels of correlated noise. The right panel of Figure \ref{fig:teffcomparison} shows how the pipeline detection efficiency depends on stellar effective temperature for a sample of stars with positive CDPP$_{\rm L}$ values. The 4000-7000~K targets still have the highest detection efficiency compared to the very cool and very hot stars, but the overall detection efficiency is reduced by the presence of the correlated noise. This is demonstrated further in Figure \ref{fig:cdppdependence}, which shows how the detection efficiency changes for a range of CDPP$_{\rm L}$ values over the full sample. For the stars with lower correlated noise (negative CDPP$_{\rm L}$ values), the behaviour is as expected, plateauing at 97\%. For the stars with higher levels of correlated noise (e.g. intrinsic stellar variability) the detection efficiency falls, plateauing at 90\%.

\begin{figure}
\centering 
\includegraphics[width=8.5cm]{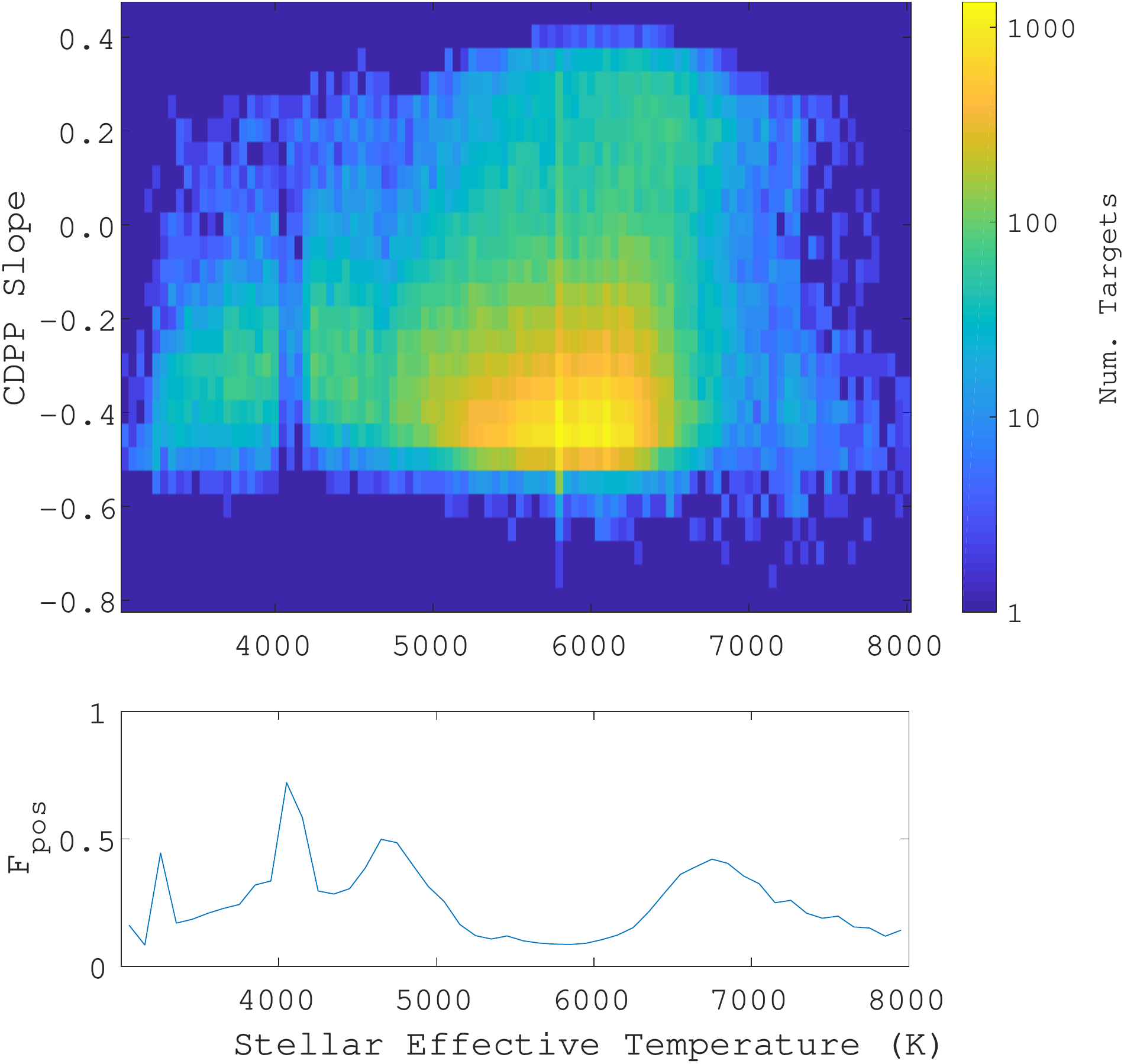}
\caption{{\it Upper}: The distribution of CDPP$_{\rm L}$ values as a function of stellar effective temperature. The bulk of the {\it Kepler} targets are well-behaved solar-like stars. {\it Lower}: The fraction of stars with positive CDPP$_{\rm L}$ values as a function of stellar effective temperature. Stars cooler than $\sim$5000~K and hotter than $\sim$6500~K are more likely to have higher levels of correlated noise in their light curves.}
\label{fig:cdppvsteff}
\end{figure}

As noted earlier, there is a correlation between the stellar temperature and the stellar noise properties. Cooler stars are more active, with more starspot and flaring activity. Figure \ref{fig:cdppvsteff} shows the distribution of CDPP$_{\rm L}$ values as a function of stellar effective temperature. 
The two peaks in the lower panel around 3200~K and 4100~K are artifacts of the available targets in those temperature regions, but it is clear that the bulk of the well-behaved (low correlated noise, negative CDPP$_{\rm L}$ values) are found from 5000--6500~K. The light curves in the right panel of Figure \ref{fig:cdppslopecomparison} are relatively well-behaved targets across the temperature range, with CDPP$_{\rm L}<-0.3$, and the light curves in the left panel are targets with much higher levels of correlated noise, with CDPP$_{\rm L}>0.1$, demonstrating the light curve morphology differences in these two populations. One way to select a stellar sample with a uniformly high detection efficiency is to restrict the selection to those targets with negative CDPP$_{\rm L}$ values; this eliminates 27,450 of the original 146,295 targets.

\begin{figure}[ht!]
\centering 
\includegraphics[width=9cm]{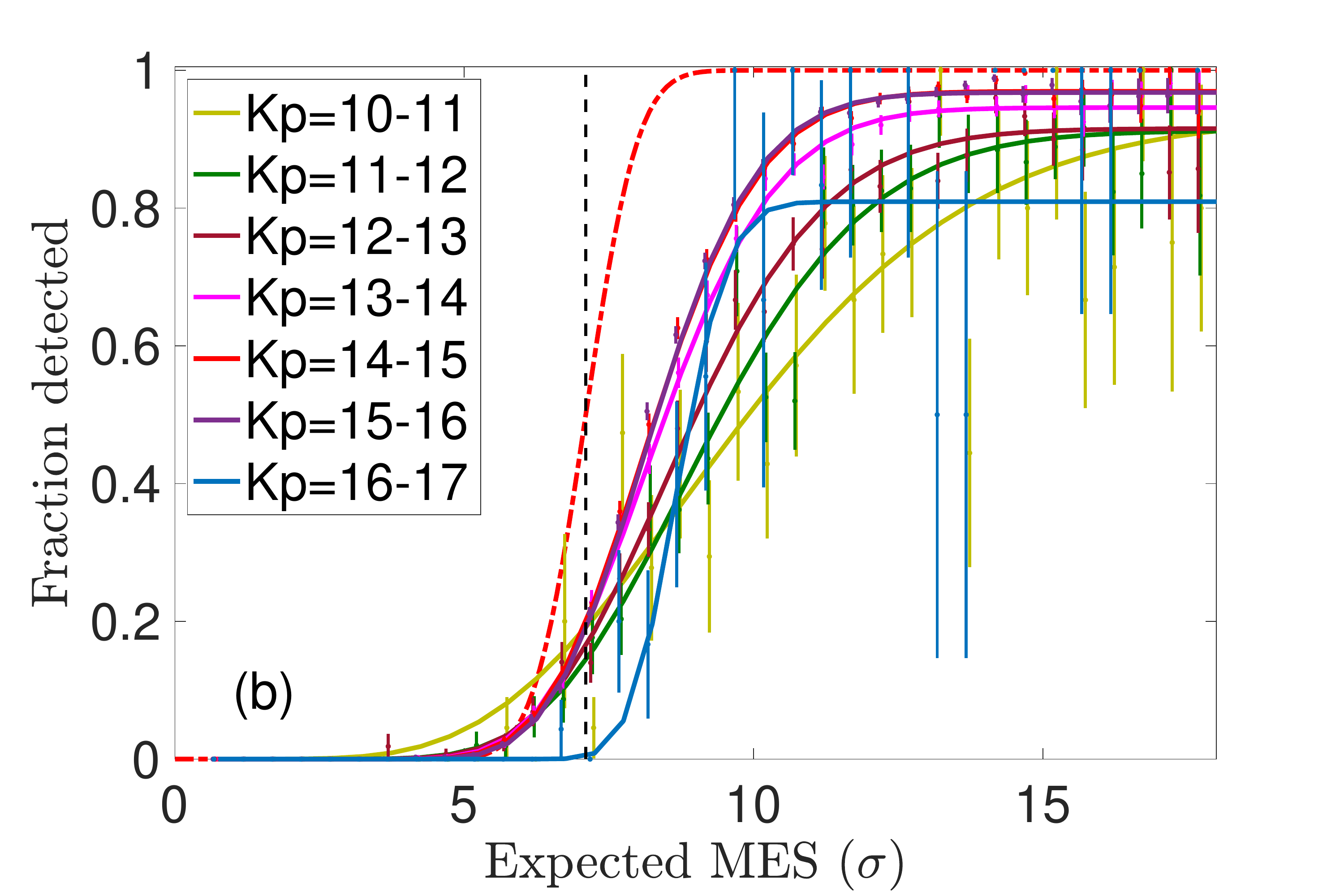}
\caption{As for Figure \ref{fig:perioddependence}, but calculated for different ranges in stellar magnitude.}
\label{fig:magdependence}
\end{figure}

In addition we examine the detection efficiency as a function of stellar magnitude. Figure \ref{fig:magdependence} shows how the detection efficiency varies as a function of magnitude. It is somewhat lower for the saturated targets ($K_p<12$), which plateau around 91--93\%, rising for the moderately bright targets ($13<Kp<16$), which plateau closer to 95--97\%), and dropping significantly for the small number of fainter targets ($K_p>16$), reaching only 81\%. The behavior at the bright end can be understood by using CDPP$_{\rm L}$ as a measure of the correlated noise in the light curves. Saturated targets typically have larger apertures to capture the bleed from the electrons that overfill the well depth. These stars have the potential for both pointing-correlated changes in flux if the apertures do not adequately capture the flux, and a greater probability of capturing flux from nearby or background targets as the area of the aperture grows. Both of these effects increase the likelihood of correlated noise; the median CDPP$_{\rm L}$ value for targets in the sample with $K_p<12$ is 0.111, whereas the same for targets with $K_p>12$ is -0.307. It is less clear why the faintest targets ($K_p>16$) have significantly reduced detection efficiency, as they have a median CDPP$_{\rm L}$ of -0.394, however there are a small number (4,579) in the sample and therefore they can be removed to select a sample with uniformly high detection efficiency.

\subsection{CCD Channel}
\label{sec:ccd}

Another variable to consider is the location of the target in the \kepler\ field of view. There are a number of CCD channels which have been identified as producing a higher number of spurious long-period TCEs due to image artifacts \citep{Thompson2018}. Section 6.7 of \citet{VanCleve2016} has additional information on the source of the artifacts. Here we examine whether target location influences the detection efficiency---in particular whether the over-abundance of long-period false positives on certain CCD channels reduces the detection efficiency of additional signals injected in the same light curve.

Since the field of view rotates every $\sim$90 days, a given group of targets (called a `sky group') will fall on four distinct CCD channels over the course of a year, symmetrically positioned around the centre of the field of view. The number of the sky group and the number of the CCD channel upon which it falls are the same in Season 2, one of the four orientations of the spacecraft around the center of the field of view. 

\begin{figure}
\centering 
\includegraphics[width=8.5cm]{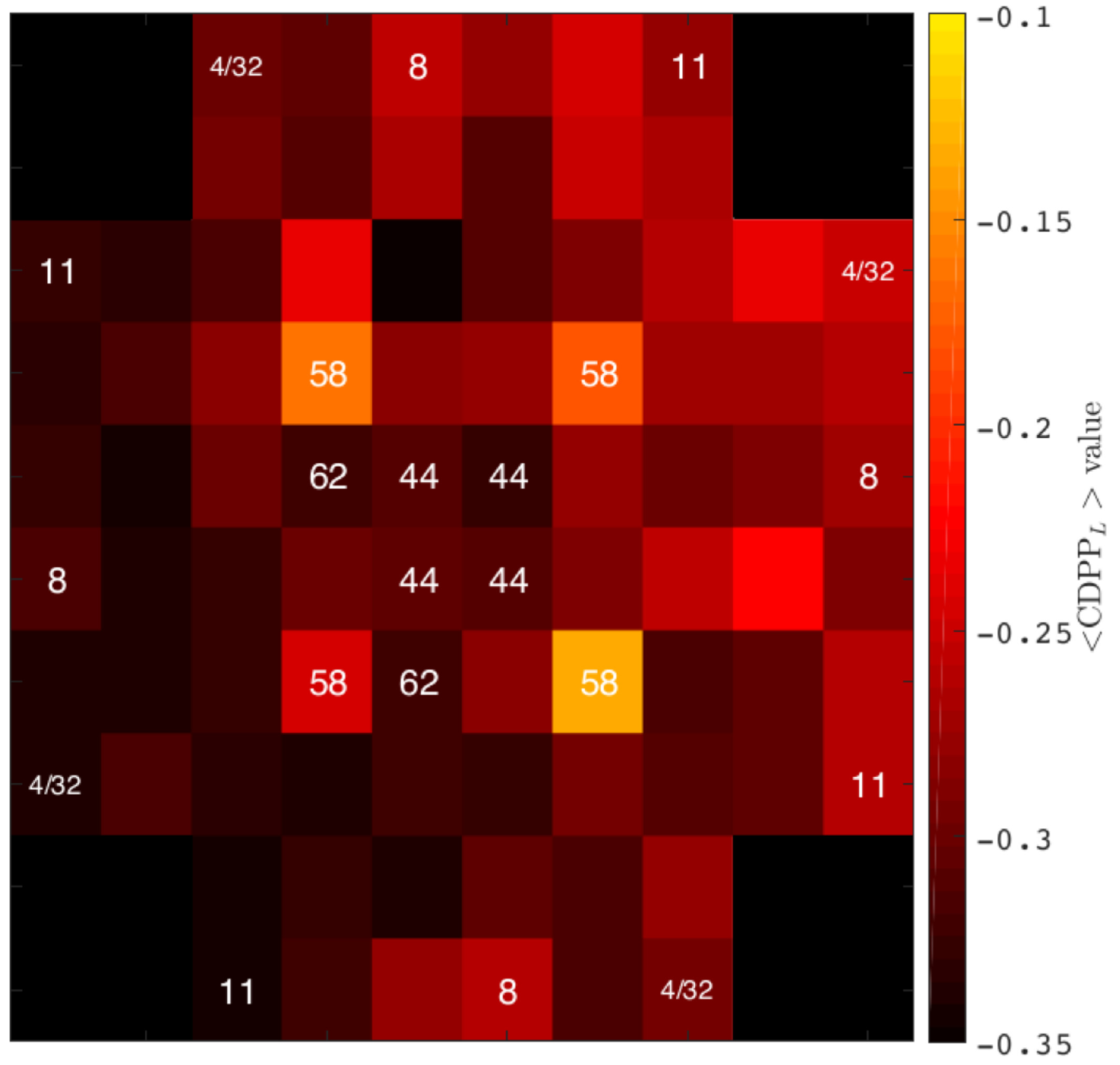}
\includegraphics[width=8.5cm]{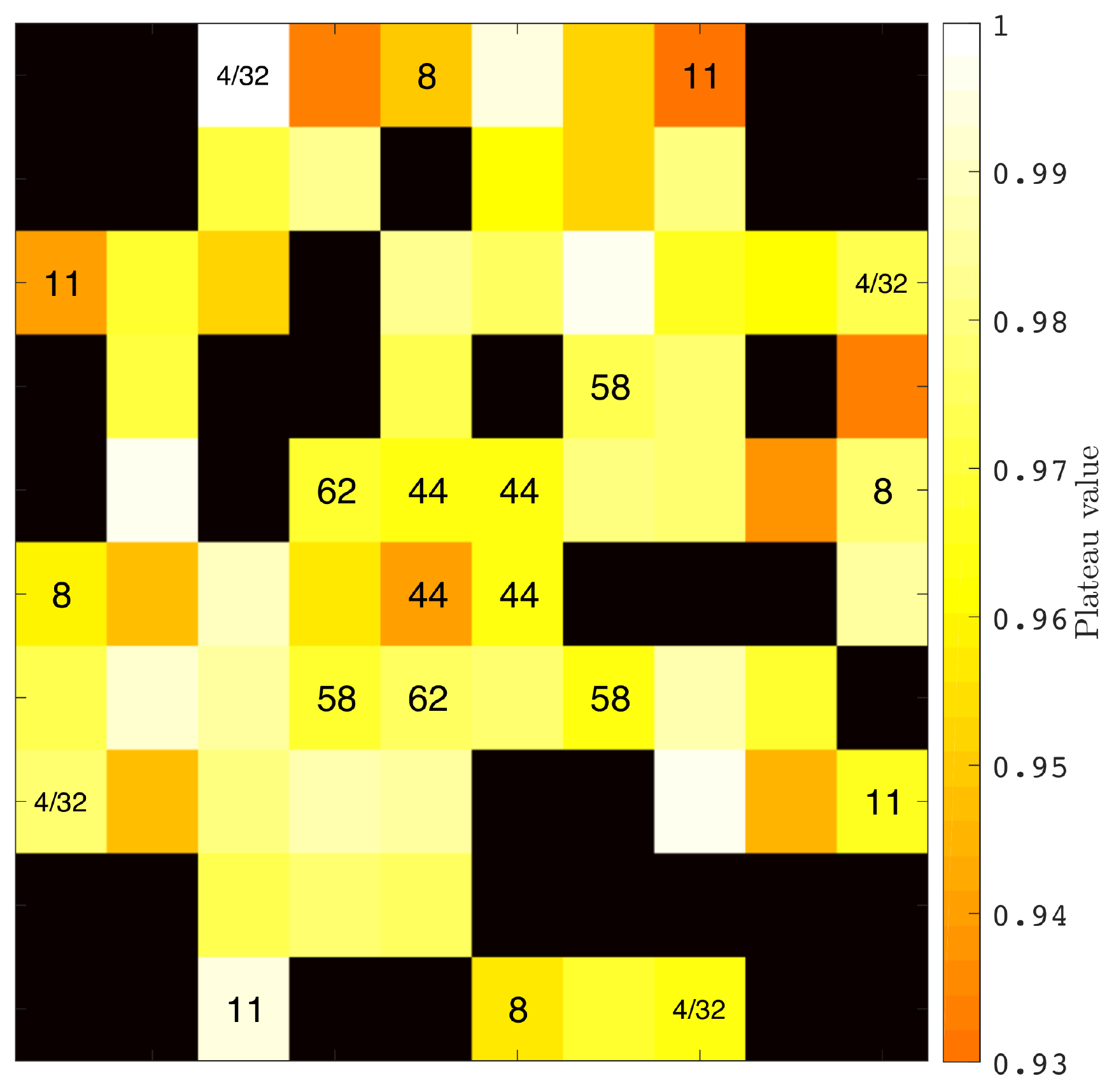}
\caption{{\it Upper:} The median CDPP$_{\rm L}$ value value for each sky group. A clear trend across the field of view is evident, tracing the Galactic latitude and worsening with higher crowding at lower Galactic latitudes. One particularly poorly performing set of sky groups which fall on CCD channel 58 during the year are labelled. {\it Lower:} The detection efficiency plateau $c$ as a function of sky group for typically well-behaved targets. Nine out of ten of the worst performing channels are in the outer ring of modules; see text for additional detail.}
\label{fig:skygroup}
\end{figure}

First we examine how the CDPP$_{\rm L}$ value varies across the field, since Section \ref{sec:stellartype} establishes the dependence of the detection efficiency on this value. The upper panel of Figure \ref{fig:skygroup} shows the median CDPP$_{\rm L}$ value across the field of view. The large black squares in the corners are the Fine Guidance Sensor CCDs, which were not used in the planet search. 
There are several notable features in the distribution of the median CDPP$_{\rm L}$ value. First, there is an underlying trend for increasing (worsening) CDPP$_{\rm L}$ values from the lower left to the upper right, which is correlated with decreasing Galactic latitude. With the relatively large \kepler\ pixels (4$^{\prime\prime}$/pixel), crowding becomes an increasing problem closer to the Galactic plane. This can increase the correlated noise measured across the CCD channel due to the increased likelihood of additional light from, e.g. background variable stars and eclipsing binaries. Four of the five channels with the highest (worst) CDPP$_{\rm L}$ values are visited by the sky groups that fall on CCD channel 58 during one of the four observing orientations. Channel 58 is one of the channels most strongly affected by the image artifacts described earlier, which is manifested here in the way the median CDPP$_{\rm L}$ value captures the correlated noise across the channel.

In the lower panel of Figure \ref{fig:skygroup} we show the distribution of the plateau values ($c$ from Equation \ref{eq:gamma}) across the field of view, for transit durations shorter than 15 hours, CDPP$_{\rm L}<0$, and at least four transits. By restricting the injections to those shown previously to have the highest detection efficiency, we can examine any remaining influence of the CCD channels on the detection efficiency. There is a weak correlation in Figure \ref{fig:skygroup} between the plateau value and the ring of best focus around the center of the focal plane \citep[see Fig. 17 of][]{VanCleve2009}---9 out of 10 channels with the lowest detection efficiencies ($<$95\%) are in the outer ring of modules (a module being a set of four CCD channels under the same field flattening lens), including the worst-performing channel 11. The CCDs with the best focus fall in a symmetric ring with a radius of $\sim2--3^{\circ}$ around the center of the focal plane (the diameter of the full focal plane is $10^{\circ}$). Therefore, any targets that fall on these CCDs will experience less contamination from nearby targets---contamination which could introduce correlated noise and reduce the detection efficiency. We also examined the intra-CCD variation in the plateau values; the standard deviation varies from 0.11 to 0.23, with the same dependence across the field of view as in the upper panel of Figure \ref{fig:skygroup}.

We have labelled some of the sky groups which fall on known problematic channels for at least one orientation of the spacecraft during the year. 
Channels 44, 58, and 62 show the largest numbers of spurious TCEs caused by image artifacts, however the sky groups that fall on these channels do not show significantly decreased detection efficiency compared to the remainder of the channels. We conclude that the detection efficiency is not degraded in these channels with many spurious long-period TCEs, which is discussed further below. 

\section{Discussion}
\label{sec:discussion}


The results presented here examine, extend, and quantify the previous indications that there are various parameters which influence the detection efficiency of the \kepler\ pipeline. For occurrence rate calculations, this implies that with careful target selection, one could increase the completeness (lower the false negative rate) of a planet candidate catalog. The reliability of such a catalog would need to be re-calculated, which is beyond the scope of this paper. For the parameters examined here, we find that most of the detection efficiency differences (stellar temperature, stellar magnitude, position on the field of view) can be captured by the dependence on the correlated noise in the light curve as summarized by CDPP$_{\rm L}$. Therefore, one can construct a target sample with a high and well-characterized completeness by removing 37,804 targets with positive CDPP$_{\rm L}$ values from the 198,709 targets searched to produced the final DR25 catalog.

One surprise was the similarity of the detection efficiency for CCDs with known correlated noise issues caused by image artifacts and of those without. As noted in \citet{Thompson2018}, CCDs strongly affected by image artifacts produce a large over-abundance of weakly detected threshold crossing events with periods 300-500 days. A previous finding by \citet{Zink2019b} had noted that the presence of multiple signals in the same light curve could degrade the detection efficiency. Therefore, it seemed likely that transit signals injected into targets on these CCDs may suffer reduced detection efficiency. However on closer investigation we find that this is not the case. The TPS module of the \kepler\ pipeline searches over a grid of period, epoch, and transit duration, and finds the combination which produces the strongest detection as measured by the Multiple Event Statistic (MES). For a light curve with multiple transiting signals, this will typically be the signal with the shortest orbital period. If this signal passes a series of vetoes, it is then removed from the light curve, which is then iteratively re-searched for additional signals which pass the MES detection threshold. This excision of data reduces the effective window function of the remaining data, and also affects the behavior of the harmonic fitter which removes sinusoidal trends in the light curve. Both of these have the effect of decreasing the detection efficiency for additional signals in the light curve. 

For this experiment, injected signals were generated uniformly in period between 0.5--500 days. The majority of these signals therefore have shorter periods than the spurious signals generated by the image artifacts, and therefore their detection efficiency seems to be largely unaffected by those longer period image artifact signals, as the injected signals are detected first and removed. Analysing the longer period ($>400$ days) injections separately, we still see similar detection efficiency between the channels strongly affected by image artifacts and those that are not. Therefore, the impact of removing long period image artifact signals on the detection efficiency of additional long period signals in the light curve seems minimal, due to the fact that removal of long period image artifact signals from the light curve removes many fewer observation points than removal of shorter period signals, creating a much smaller reduction in the subsequent window function.

\begin{figure}
\centering 
\includegraphics[width=8.5cm]{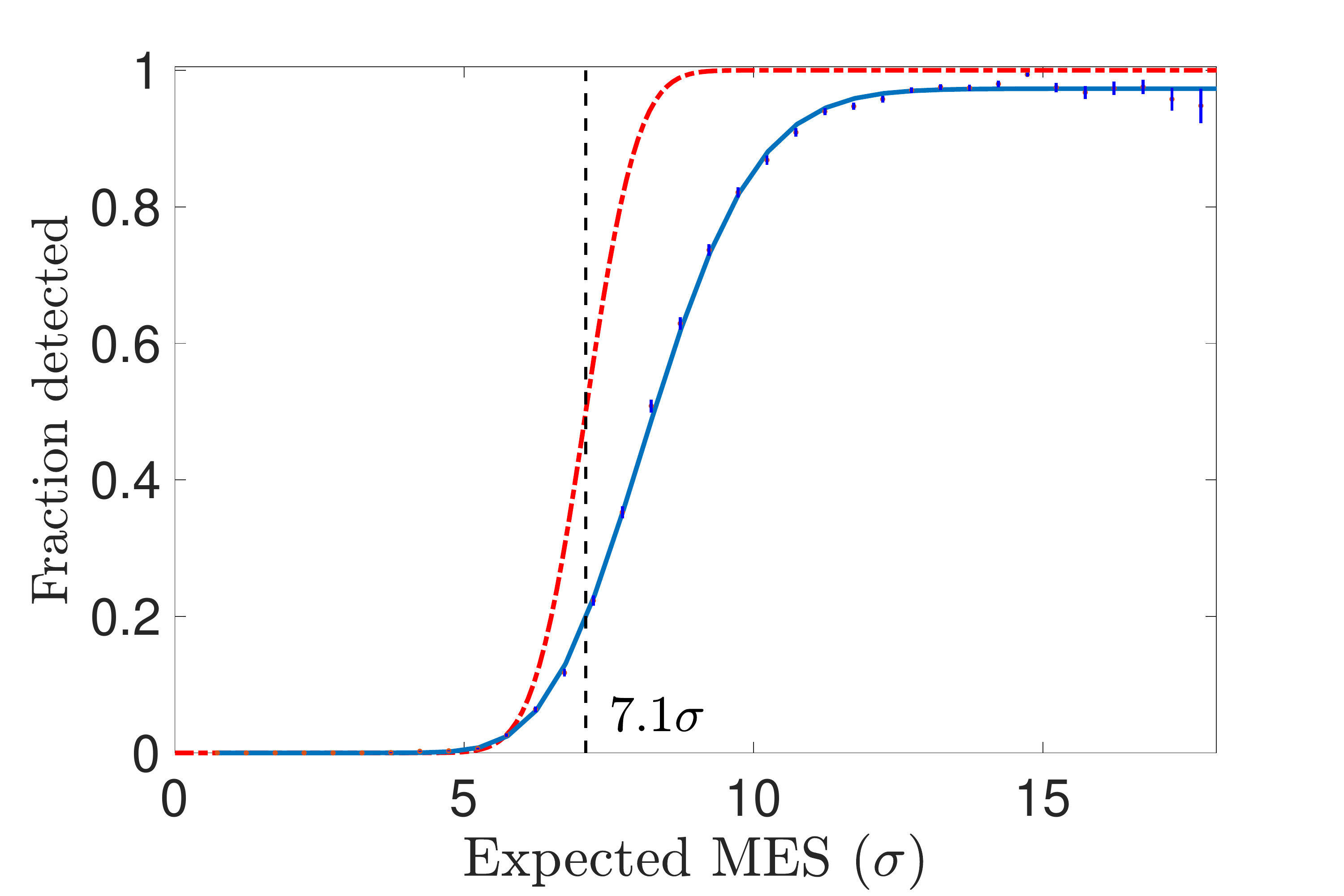}
\caption{As for Figure \ref{fig:perioddependence}, but calculated for the set of stars with CDPP$_{\rm L} \le 0$, for injected signals with four or more transits, orbital periods $<350$ days, and transit durations $<15$ hours.}
\label{fig:finalcompleteness}
\end{figure}

However, independent of the location on the field of view, we do show that the most complete planet candidate catalog is that which is confined to signals with periods $<350$ days. None of which is to say that one should not or cannot perform occurrence rate calculations for targets or signals outside of the bounds enumerated in this work, but that one would have to calculate and apply the appropriate completeness correction for the desired sample. For the target, sample defined above, with CDPP$_{\rm L}<0$, the detection efficiency for injections with four or more transits, with injected transit durations shorter than 15 hours, and with orbital periods shorter than 350 days, is shown in Figure \ref{fig:finalcompleteness}. The final best fit using Equation \ref{eq:gamma} is $\alpha=33.54$, $\beta=0.2478$, and a plateau of $c=97.31$\%.

As discussed in Section \ref{sec:experiment}, due to limitations on available resources the transit injection experiments were performed orthogonally---the wide-and-shallow pixel-level transit injections, with one injected signal per target, and the narrow-and-deep flux-level transit injections, with many thousands of signals injected into a smaller number of targets. A limitation of the pixel-level transit injections described in this work is that it averages over any fine structure in the response of the detection efficiency to features of the targets or the instrument, and provides only a description of the ensemble behavior. The flux-level transit injections were crucial, for instance, in identifying the role of the CDPP slope in discriminating between well- and poorly-behaved targets. However due to the relatively small number of targets probed by the flux-level transit injections, there may remain parameters to which the detection efficiency of the \kepler\ pipeline is sensitive that we have not yet identified. All of the light curves with simulated signals from the pixel- and flux-level transit injection experiments are available at the NASA Exoplanet Archive for further scrutiny by the community\footnote{\url{https://exoplanetarchive.ipac.caltech.edu/docs/KeplerSimulated.html}, DOI 10.26133/NEA14}.

\section{Conclusions}
\label{sec:conclusion}

This concludes the final \kepler\ project analysis of the detection efficiency of the \kepler\ pipeline. The performance of the SOC version 9.3 pipeline \citep{Twicken2016} in producing the Data Release 25 planet candidate catalog \citep{Thompson2018} was found to be a return to high detection efficiency after a moderate decrease in SOC version 9.2 \citep{Christiansen2016}. The dependence of the detection efficiency on the properties of both the targets and instrument were explored in some detail---the pipeline was found to have the highest detection efficiency for FGK stars ($4000\le T_{\rm eff} \le 7000$~K) with well-behaved noise properties. CCD channels with poor focus were found to have decreased detection efficiency, as were signals with periods longer than 300 days. 

The fact that the response of the \kepler\ pipeline varied so strongly in different target and instrument parameter space speaks to the importance of transit injection experiments like the work described here. Analogous studies using data from missions like \emph{K2} and \emph{TESS}, which have the opportunity to extend the occurrence rate results of \kepler\ beyond main sequence FGK stars, will similarly need to quantify the completeness and reliability of their resulting planet candidate catalogs to facilitate the generation of robust occurrence rates.

\acknowledgments

We thank the anonymous referee for thoughtful comments and questions that improved the manuscript. Funding for the \kepler\ Discovery Mission is provided by NASA's Science Mission Directorate. These data products were generated by the Kepler Mission science pipeline through the efforts of the Kepler Science Operations Center and Science Office. This research has made use of the NASA Exoplanet Archive, which is operated by the California Institute of Technology, under contract with the National Aeronautics and Space Administration under the Exoplanet Exploration Program.



Facilities: \facility{Kepler}.


\clearpage

\appendix
\label{app:pipeline}

The end-of-mission version of the SOC pipeline has been described in considerable detail in \citet{Jenkins2017a} and chapters therein; see Fig. 1 of that document for an overview. The code itself is also available online\footnote{\url{https://github.com/nasa/kepler-pipeline}}. Initially, the raw pixels are calibrated by the CAL module \citep{Clarke2017}, including corrections for bias, gain, non-linearity, flat-field and local detector electronics effects (overshoot and undershoot). There is also a correction for the smearing of the image that results from the fact that \emph{Kepler} operates without a shutter. In version 9.3, the bias correction was updated from a static two-dimensional correction to fully dynamic two-dimensional correction. This allowed the calibration to capture changes due to drifts in the bias values, such as those caused by temperature changes or crosstalk in the CCD electronics.

The calibrated pixels are then used to generate simple aperture photometry (SAP) time series in the PA module \citep{Morris2017}. Due to its extremely stable and precise pointing, the aperture photometry is generated by summing over whole, discrete pixels (as compared to fractions thereof). Prior to this version of the pipeline, the pixels chosen for inclusion in the photometric aperture were calculated by predicting the signal-to-noise (SNR) of the flux contribution of the target star to each pixel by using a model of the CCD, the pixel response function and the Kepler Input Catalog \citep{Brown2011}. In version 9.3, the procedure was updated to use the calibrated pixels themselves to calculate the SNR \citep{Smith2017a}.

The time series are then corrected for systematic noise components in the Presearch Data Conditioning (PDC) module \citep{Smith2017b}. In version 9.3, PDC was updated to include `spike' basis vectors that corrected for individual observations which triggered an inordinate number of spurious transit detections across multiple targets. In addition, the previous decomposition of the Bayesian Maximum A Posteriori (MAP) correction \citep{Smith2012} into multiple timescales was extended to include the shortest (1.5-hr) timescale. As noted by \citet{Twicken2016}, the improvements in version 9.3 to the generation and treatment of the time series decreased their noise, as measured by the Combined Differential Photometric Precision \citep{Christiansen2012}, by a few percent on average. 

The corrected time series are then searched for periodic transit-like signals by the Transiting Planet Search (TPS) module \citep{Jenkins2017b}. TPS first prepares the time series, removing harmonic features and various flavours of outliers and then applying a wavelet-based matched filter to whiten the noise (i.e. equalise the noise contributions across frequencies). The time series is then searched for periodic signals with at least three events, and a statistical significance as measured by the Multiple Event Statistic (MES) exceeding the $7.1\sigma$ threshold. For each identified Threshold Crossing Event (TCE), TPS then performs a number of additional checks to examine the robustness and uniformity of the signal, and vetoes signals which do not pass the checks. 

There were several important updates to TPS in version 9.3. These included: (i) the number of harmonic components removed in each quarter was made a function of the length of the quarter, to reduce over-fitting of signals in short quarters; (ii) the whitening was performed quarter-by-quarter instead on the time series as a whole, to compensate for discrete noise properties in each quarter; (iii) the update to the whitening algorithm necessitated an update to the long ($>2.5$-day) gap filling algorithm, using a sigmoid taper instead of a linear taper at the centre of the gap, to avoid the artifically low noise properties that were occurring in the gaps; and (iv) the rms noise calculations performed in the wavelet analysis were updated to use a non-decimated moving median absolute deviation (MAD), to more accurately represent the noise properties. All of these updates were designed to improve the sensitivity of the transit search.

Once a given time series is found to host a TCE that passes all the vetoes, it is passed to the Data Validation (DV) module \citep{Twicken2018,Li2019}. DV performs a transit fit for the first TCE, using the \citet{Mandel2002} prescription. The in-transit observations are then removed, and the subsequent time series is then sent back to TPS for additional scrutiny. This is repeated until the time series produces no more TCEs, the limit of the number of TCEs (10) is reached, or the time limit\footnote{Time limits on the NASA Pleiades super-computer are imposed for resource management reasons.} for DV to search a given time series is reached. DV then produces a suite of diagnostic tests and plots for each TCE.

The process by which TCEs were dispositioned into Kepler Objects of Interest (KOIs) and then into planet candidates (PCs) or false positives (FPs) evolved considerably over the course of the \emph{Kepler} mission. From individual decisions made by eye, to team decisions made by multiple eyes, to team decisions made using a set of metrics, to a suite of algorithms dubbed the `Robovetter' automatically evaluating that set of metrics. Taking the people out of the process was the most important step towards quantifying the detection efficiency of the process, and the prime motivation towards development of the Robovetter.

For DR25, the final list of TCEs and diagnostics is passed to the Robovetter. The details are provided in \citet{Coughlin2016} and updated in \citet{Thompson2018}; see Fig 4. of the latter for a schematic overview. Tab. 3 of \citet{Thompson2018} provides the full suite of tests performed by the Robovetter, but in summary, in order for a TCE to be promoted to a planet candidate, it must satisfy a number of criteria, the most important of which are: 
\\
\begin{enumerate}
    \item Not have an ephemeris which matches that of a previously identified TCE in any light curve, including the light curve being analysed;
    \item Not have a secondary eclipse inconsistent with a planetary origin;
    \item Not have statistically significant depth changes between the odd-numbered events and the even-numbered events (indicating an eclipsing binary system);
    \item Not have a significantly v-shaped folded transit signal (also indicating an eclipsing binary system);
    \item Have consistent depths for all measured transits (such that the folded signal strength is not dominated by one deeper event);
    \item Be unique and statistically significant when compared to the correlated noise properties of the light curve when folded at the period of the TCE; and
    \item Comprise at least three transits that have all individually passed a battery of additional tests interrogating their shape, coverage, and whether they fall during times which produce an inordinately high number of (likely spurious) signals.
\end{enumerate}

Criteria 1 (identifying ephemeris matches) eliminates $\sim$0.05\% of injected planets, Criteria 2--5 (identifying stellar eclipses) eliminate $\sim$1.3\% of injected planets, and Criteria 6--7 (identifying non-transit-like events) eliminate $\sim$12\% of injected planets; see \citet{Thompson2018} for additional details.

In the final mission-supported run of the Robovetter, the algorithms were tuned to maximise the reliability of the catalog for a given minimum completeness of shallow signals at long periods. This necessarily resulted in a sacrifice in the completeness of the catalog, or the number of the true positives promoted by the Robovetter. The goal of the transit injection and recovery experiment described in this work is to quantify the fraction of true positives that are lost as a result of this fine tuning. In DR25, the final run of the Robovetter resulted in 8,054 TCEs being classified as Kepler Objects of Interest (KOIs), and 4,034 of those as planet candidates (PCs).





\clearpage

\bibliographystyle{apj}
\bibliography{refs}

\begin{thebibliography}{}
\expandafter\ifx\csname natexlab\endcsname\relax\def\natexlab#1{#1}\fi

\bibitem[{{Akeson} {et~al.}(2013){Akeson}, {Chen}, {Ciardi}, {Crane}, {Good},
  {Harbut}, {Jackson}, {Kane}, {Laity}, {Leifer}, {Lynn}, {McElroy}, {Papin},
  {Plavchan}, {Ram{\'{\i}}rez}, {Rey}, {von Braun}, {Wittman}, {Abajian},
  {Ali}, {Beichman}, {Beekley}, {Berriman}, {Berukoff}, {Bryden}, {Chan},
  {Groom}, {Lau}, {Payne}, {Regelson}, {Saucedo}, {Schmitz}, {Stauffer},
  {Wyatt}, \& {Zhang}}]{Akeson2013}
{Akeson}, R.~L., {Chen}, X., {Ciardi}, D., {et~al.} 2013, \pasp, 125, 989

\bibitem[{{Berger} {et~al.}(2018){Berger}, {Huber}, {Gaidos}, \& {van
  Saders}}]{berger18}
{Berger}, T.~A., {Huber}, D., {Gaidos}, E., \& {van Saders}, J.~L. 2018, \apj,
  866, 99

\bibitem[{{Brown} {et~al.}(2011){Brown}, {Latham}, {Everett}, \&
  {Esquerdo}}]{Brown2011}
{Brown}, T.~M., {Latham}, D.~W., {Everett}, M.~E., \& {Esquerdo}, G.~A. 2011,
  \aj, 142, 112

\bibitem[{{Burke} \& {Catanzarite}(2017{\natexlab{a}})}]{BurkeJCat2017b}
{Burke}, C.~J., \& {Catanzarite}, J. 2017{\natexlab{a}}, NASA Exoplanet
  Archive, KSCI-19111-002

\bibitem[{{Burke} \& {Catanzarite}(2017{\natexlab{b}})}]{BurkeJCat2017a}
---. 2017{\natexlab{b}}, NASA Exoplanet Archive, KSCI-19109-002

\bibitem[{{Burke} \& {Catanzarite}(2017{\natexlab{c}})}]{Burke2017window}
---. 2017{\natexlab{c}}, {Planet Detection Metrics: Window and One-Sigma Depth
  Functions for Data Release 25}, Tech. rep.

\bibitem[{{Christiansen}(2017)}]{Christiansen2017}
{Christiansen}, J.~L. 2017, NASA Exoplanet Archive, KSCI-19110-001

\bibitem[{{Christiansen} {et~al.}(2012){Christiansen}, {Jenkins}, {Caldwell},
  {Burke}, {Tenenbaum}, {Seader}, {Thompson}, {Barclay}, {Clarke}, {Li},
  {Smith}, {Stumpe}, {Twicken}, \& {Van Cleve}}]{Christiansen2012}
{Christiansen}, J.~L., {Jenkins}, J.~M., {Caldwell}, D.~A., {et~al.} 2012,
  \pasp, 124, 1279

\bibitem[{{Christiansen} {et~al.}(2013){Christiansen}, {Clarke}, {Burke},
  {Jenkins}, {Barclay}, {Ford}, {Haas}, {Sabale}, {Seader}, {Claiborne Smith},
  {Tenenbaum}, {Twicken}, {Kamal Uddin}, \& {Thompson}}]{Christiansen2013}
{Christiansen}, J.~L., {Clarke}, B.~D., {Burke}, C.~J., {et~al.} 2013,
  Astrophysical Journal Supplement, 207, 35

\bibitem[{{Christiansen} {et~al.}(2015){Christiansen}, {Clarke}, {Burke},
  {Seader}, {Jenkins}, {Twicken}, {Catanzarite}, {Smith}, {Batalha}, {Haas},
  {Thompson}, {Campbell}, {Sabale}, \& {Kamal Uddin}}]{Christiansen2015}
---. 2015, Astrophysical Journal, 810, 95

\bibitem[{{Christiansen} {et~al.}(2016){Christiansen}, {Clarke}, {Burke},
  {Jenkins}, {Bryson}, {Coughlin}, {Mullally}, {Thompson}, {Twicken},
  {Batalha}, {Haas}, {Catanzarite}, {Campbell}, {Kamal Uddin}, {Zamudio},
  {Smith}, \& {Henze}}]{Christiansen2016}
---. 2016, Astrophysical Journal, 828, 99

\bibitem[{{Clarke} {et~al.}(2017){Clarke}, {Caldwell}, {Quintana},
  {Chandrasekaran}, {Twicken}, {Jenkins}, {Cote}, {McCauliff}, {Klaus},
  {Allen}, \& {Bryson}}]{Clarke2017}
{Clarke}, B.~D., {Caldwell}, D.~A., {Quintana}, E.~V., {et~al.} 2017, {Kepler
  Data Processing Handbook: Pixel Level Calibrations}, Tech. rep.

\bibitem[{{Coughlin}(2017)}]{Coughlin2017}
{Coughlin}, J.~L. 2017, Planet Detection Metrics: Robovetter Completeness and
  Effectiveness for Data Release 25 (KSCI-19114-002)

\bibitem[{Coughlin {et~al.}(2016)Coughlin, Mullally, Thompson, Rowe, Burke,
  Latham, Batalha, Ofir, Quarles, Henze, Wolfgang, Caldwell, Bryson, Shporer,
  Catanzarite, Akeson, Barclay, Borucki, Boyajian, Campbell, Christiansen,
  Girouard, Haas, Howell, Huber, Jenkins, Li, Patil-Sabale, Quintana, Ramirez,
  Seader, Smith, Tenenbaum, Twicken, \& Zamudio}]{Coughlin2016}
Coughlin, J.~L., Mullally, F., Thompson, S.~E., {et~al.} 2016, The
  Astrophysical Journal Supplement Series, 224, 12

\bibitem[{{Jenkins}(2017)}]{Jenkins2017a}
{Jenkins}, J.~M. 2017, {Kepler Data Processing Handbook: Philosophy and Scope},
  Tech. rep.

\bibitem[{{Jenkins} {et~al.}(1996){Jenkins}, {Doyle}, \&
  {Cullers}}]{Jenkins1996}
{Jenkins}, J.~M., {Doyle}, L.~R., \& {Cullers}, D.~K. 1996, \icarus, 119, 244

\bibitem[{{Jenkins} {et~al.}(2017){Jenkins}, {Tenenbaum}, {Seader}, {Burke},
  {McCauliff}, {Smith}, {Twicken}, \& {Chandrasekaran}}]{Jenkins2017b}
{Jenkins}, J.~M., {Tenenbaum}, P., {Seader}, S., {et~al.} 2017, {Kepler Data
  Processing Handbook: Transiting Planet Search}, Tech. rep.

\bibitem[{{Jenkins} {et~al.}(2010{\natexlab{a}}){Jenkins}, {Chandrasekaran},
  {McCauliff}, {Caldwell}, {Tenenbaum}, {Li}, {Klaus}, {Cote}, \&
  {Middour}}]{JenkinsSPIE2010}
{Jenkins}, J.~M., {Chandrasekaran}, H., {McCauliff}, S.~D., {et~al.}
  2010{\natexlab{a}}, Society of Photo-Optical Instrumentation Engineers (SPIE)
  Conference Series, Vol. 7740, {Transiting planet search in the Kepler
  pipeline}, 77400D

\bibitem[{{Jenkins} {et~al.}(2010{\natexlab{b}}){Jenkins}, {Chandrasekaran},
  {McCauliff}, {Caldwell}, {Tenenbaum}, {Li}, {Klaus}, {Cote}, \&
  {Middour}}]{Jenkins2010}
{Jenkins}, J.~M., {Chandrasekaran}, H., {McCauliff}, S.~D., {et~al.}
  2010{\natexlab{b}}, in Society of Photo-Optical Instrumentation Engineers
  (SPIE) Conference Series, Vol. 7740, \procspie, 77400D

\bibitem[{{Li} {et~al.}(2019){Li}, {Tenenbaum}, {Twicken}, {Burke}, {Jenkins},
  {Quintana}, {Rowe}, \& {Seader}}]{Li2019}
{Li}, J., {Tenenbaum}, P., {Twicken}, J.~D., {et~al.} 2019, \pasp, 131, 024506

\bibitem[{{Mandel} \& {Agol}(2002)}]{Mandel2002}
{Mandel}, K., \& {Agol}, E. 2002, \apjl, 580, L171

\bibitem[{{Mathur} {et~al.}(2017){Mathur}, {Huber}, {Batalha}, {Ciardi},
  {Bastien}, {Bieryla}, {Buchhave}, {Cochran}, {Endl}, {Esquerdo}, {Furlan},
  {Howard}, {Howell}, {Isaacson}, {Latham}, {MacQueen}, \&
  {Silva}}]{Mathur2017ApJS}
{Mathur}, S., {Huber}, D., {Batalha}, N.~M., {et~al.} 2017, \apjs, 229, 30

\bibitem[{{Morris} {et~al.}(2017){Morris}, {Twicken}, {Smith}, {Clarke},
  {Jenkins}, {Bryson}, {Girouard}, \& {Klaus}}]{Morris2017}
{Morris}, R.~L., {Twicken}, J.~D., {Smith}, J.~C., {et~al.} 2017, {Kepler Data
  Processing Handbook: Photometric Analysis}, Tech. rep.

\bibitem[{{Mullally} {et~al.}(2015){Mullally}, {Coughlin}, {Thompson}, {Rowe},
  {Burke}, {Latham}, {Batalha}, {Bryson}, {Christiansen}, {Henze}, {Ofir},
  {Quarles}, {Shporer}, {Van Eylen}, {Van Laerhoven}, {Shah}, {Wolfgang},
  {Chaplin}, {Xie}, {Akeson}, {Argabright}, {Bachtell}, {Barclay}, {Borucki},
  {Caldwell}, {Campbell}, {Catanzarite}, {Cochran}, {Duren}, {Fleming},
  {Fraquelli}, {Girouard}, {Haas}, {He{\l}miniak}, {Howell}, {Huber}, {Larson},
  {Gautier}, {Jenkins}, {Li}, {Lissauer}, {McArthur}, {Miller}, {Morris},
  {Patil-Sabale}, {Plavchan}, {Putnam}, {Quintana}, {Ramirez}, {Silva Aguirre},
  {Seader}, {Smith}, {Steffen}, {Stewart}, {Stober}, {Still}, {Tenenbaum},
  {Troeltzsch}, {Twicken}, \& {Zamudio}}]{Mullally2015}
{Mullally}, F., {Coughlin}, J.~L., {Thompson}, S.~E., {et~al.} 2015, \apjs,
  217, 31

\bibitem[{{Rowe} {et~al.}(2015){Rowe}, {Coughlin}, {Antoci}, {Barclay},
  {Batalha}, {Borucki}, {Burke}, {Bryson}, {Caldwell}, {Campbell},
  {Catanzarite}, {Christiansen}, {Cochran}, {Gilliland}, {Girouard}, {Haas},
  {He{\l}miniak}, {Henze}, {Hoffman}, {Howell}, {Huber}, {Hunter},
  {Jang-Condell}, {Jenkins}, {Klaus}, {Latham}, {Li}, {Lissauer}, {McCauliff},
  {Morris}, {Mullally}, {Ofir}, {Quarles}, {Quintana}, {Sabale}, {Seader},
  {Shporer}, {Smith}, {Steffen}, {Still}, {Tenenbaum}, {Thompson}, {Twicken},
  {Van Laerhoven}, {Wolfgang}, \& {Zamudio}}]{Rowe2015}
{Rowe}, J.~F., {Coughlin}, J.~L., {Antoci}, V., {et~al.} 2015, \apjs, 217, 16

\bibitem[{{Smith} {et~al.}(2017{\natexlab{a}}){Smith}, {Morris}, {Jenkins},
  {Bryson}, {Caldwell}, \& {Girouard}}]{Smith2017a}
{Smith}, J.~C., {Morris}, R.~L., {Jenkins}, J.~M., {et~al.} 2017{\natexlab{a}},
  {Kepler Data Processing Handbook: Finding Optimal Apertures in Kepler Data},
  Tech. rep.

\bibitem[{{Smith} {et~al.}(2012){Smith}, {Stumpe}, {Van Cleve}, {Jenkins},
  {Barclay}, {Fanelli}, {Girouard}, {Kolodziejczak}, {McCauliff}, {Morris}, \&
  {Twicken}}]{Smith2012}
{Smith}, J.~C., {Stumpe}, M.~C., {Van Cleve}, J.~E., {et~al.} 2012, \pasp, 124,
  1000

\bibitem[{{Smith} {et~al.}(2017{\natexlab{b}}){Smith}, {Stumpe}, {Jenkins},
  {Van Cleve}, {Girouard}, {Kolodziejczak}, {McCauliff}, {Morris}, \&
  {Twicken}}]{Smith2017b}
{Smith}, J.~C., {Stumpe}, M.~C., {Jenkins}, J.~M., {et~al.} 2017{\natexlab{b}},
  {Kepler Data Processing Handbook: Presearch Data Conditioning}, Tech. rep.

\bibitem[{{Thompson} {et~al.}(2018){Thompson}, {Coughlin}, {Hoffman},
  {Mullally}, {Christiansen}, {Burke}, {Bryson}, {Batalha}, {Haas},
  {Catanzarite}, {Rowe}, {Barentsen}, {Caldwell}, {Clarke}, {Jenkins}, {Li},
  {Latham}, {Lissauer}, {Mathur}, {Morris}, {Seader}, {Smith}, {Klaus},
  {Twicken}, {Van Cleve}, {Wohler}, {Akeson}, {Ciardi}, {Cochran}, {Henze},
  {Howell}, {Huber}, {Pr{\v s}a}, {Ram{\'{\i}}rez}, {Morton}, {Barclay},
  {Campbell}, {Chaplin}, {Charbonneau}, {Christensen-Dalsgaard}, {Dotson},
  {Doyle}, {Dunham}, {Dupree}, {Ford}, {Geary}, {Girouard}, {Isaacson},
  {Kjeldsen}, {Quintana}, {Ragozzine}, {Shabram}, {Shporer}, {Silva Aguirre},
  {Steffen}, {Still}, {Tenenbaum}, {Welsh}, {Wolfgang}, {Zamudio}, {Koch}, \&
  {Borucki}}]{Thompson2018}
{Thompson}, S.~E., {Coughlin}, J.~L., {Hoffman}, K., {et~al.} 2018,
  Astrophysical Journal Supplement, 235, 38

\bibitem[{{Twicken} {et~al.}(2016){Twicken}, {Jenkins}, {Seader}, {Tenenbaum},
  {Smith}, {Brownston}, {Burke}, {Catanzarite}, {Clarke}, {Cote}, {Girouard},
  {Klaus}, {Li}, {McCauliff}, {Morris}, {Wohler}, {Campbell}, {Kamal Uddin},
  {Zamudio}, {Sabale}, {Bryson}, {Caldwell}, {Christiansen}, {Coughlin},
  {Haas}, {Henze}, {Sanderfer}, \& {Thompson}}]{Twicken2016}
{Twicken}, J.~D., {Jenkins}, J.~M., {Seader}, S.~E., {et~al.} 2016,
  Astrophysical Journal, 152, 158

\bibitem[{{Twicken} {et~al.}(2018){Twicken}, {Catanzarite}, {Clarke},
  {Girouard}, {Jenkins}, {Klaus}, {Li}, {McCauliff}, {Seader}, {Tenenbaum},
  {Wohler}, {Bryson}, {Burke}, {Caldwell}, {Haas}, {Henze}, \&
  {Sanderfer}}]{Twicken2018}
{Twicken}, J.~D., {Catanzarite}, J.~H., {Clarke}, B.~D., {et~al.} 2018, \pasp,
  130, 064502

\bibitem[{{Van Cleve} \& {Caldwell}(2009)}]{VanCleve2009}
{Van Cleve}, J.~E., \& {Caldwell}, D.~A. 2009, MAST Archive, KSCI-19033-001

\bibitem[{{Van Cleve} {et~al.}(2009){Van Cleve}, Christiansen, Jenkins,
  Caldwell, Barclay, Bryson, Burke, Campbell, Catanzarite, Clarke, Coughlin,
  Girouard, Haas, Klaus, Kolodziejczak, Li, McCauliff, Morris, Mullally,
  Quintana, Rowe, Sabale, Seader, Smith, Still, Tenenbaum, Thompson, Twicken,
  Uddin, \& Zamudio}]{VanCleve2016}
{Van Cleve}, J.~E., Christiansen, J.~L., Jenkins, J.~M., {et~al.} 2009, MAST
  Archive, KSCI-19040-005

\bibitem[{{Zink} \& {Hansen}(2019)}]{Zink2019b}
{Zink}, J.~K., \& {Hansen}, B. M.~S. 2019, \mnras, 487, 246

\end{thebibliography}

\clearpage



\clearpage








\clearpage


\end{document}